\documentclass[aps,prd,reprint,longbibliography,nofootinbib,superscriptaddress]{revtex4-2}

\usepackage{lmodern}
\usepackage[T1]{fontenc}
\usepackage{amsmath}
\usepackage{amssymb}

\usepackage{url}
\usepackage{multirow}

\usepackage[nodayofweek]{datetime}
\usepackage{enumerate}
\usepackage{xspace}
\usepackage{xcolor}
\usepackage{graphicx}
\usepackage[pdftex,hidelinks]{hyperref}
\usepackage{bookmark}

\usepackage{tabularx}
\usepackage{booktabs}

\usepackage{float}

\usepackage[capitalise]{cleveref}

\newcommand{\diagramscale}{0.7}
\newcommand{\plotscale}{0.39}

\newcommand{\pippin}{PIPPIN\xspace}
\newcommand{\pippinfull}{Particles Into Particles with Permutation Invariant Network\xspace}
\newcommand{\pippintitle}{\pippin: Generating variable length full events from partons\xspace}
\newcommand{\geant}{\textsc{Geant4}\xspace}
\newcommand{\madgraph}{\textsc{MadGraph5}\xspace}
\newcommand{\pythia}{\textsc{Pythia}\xspace}
\newcommand{\delphes}{\textsc{Delphes}\xspace}

\newcommand{\pt}{\ensuremath{{p}_\mathrm{T}}\xspace}

\makeatletter\def\frontmatter@affiliationfont{\it\footnotesize}\makeatother

\begin{document}

\title{\pippintitle}

\author{Guillaume Quétant}
\email{guillaume.quetant@unige.ch}
\affiliation{Département de Physique Nucléaire et Corpusculaire, University of Geneva, Switzerland}
\affiliation{Département d'Informatique, University of Geneva, Switzerland}

\author{John Andrew Raine}
\affiliation{Département de Physique Nucléaire et Corpusculaire, University of Geneva, Switzerland}

\author{Matthew Leigh}
\affiliation{Département de Physique Nucléaire et Corpusculaire, University of Geneva, Switzerland}

\author{Debajyoti Sengupta}
\affiliation{Département de Physique Nucléaire et Corpusculaire, University of Geneva, Switzerland}

\author{Tobias Golling}
\affiliation{Département de Physique Nucléaire et Corpusculaire, University of Geneva, Switzerland}

\begin{abstract}
    This paper presents a novel approach for directly generating full events at detector-level from parton-level information, leveraging cutting-edge machine learning techniques.
To address the challenge of multiplicity variations between parton and reconstructed object spaces, we employ transformers, score-based models and normalizing flows.
Our method tackles the inherent complexities of the stochastic transition between these two spaces and achieves remarkably accurate results.
The combination of innovative techniques and the achieved accuracy demonstrates the potential of our approach in advancing the field and opens avenues for further exploration.
This research contributes to the ongoing efforts in high-energy physics and generative modelling, providing a promising direction for enhanced precision in fast detector simulation.

\end{abstract}

\maketitle

\section{Introduction}
In the realm of high-energy physics, the simulation of particle collisions is a crucial tool for the downstream analysis of the huge amount of data produced by collider experiments.
The classical generation of these simulated events is a complex task.
By far the most computationally taxing subtask is the propagation of particles through the detectors, accounting for the interactions with the detector material and modelling the secondary radiation showers this produces.
This process is usually performed using Monte Carlo (MC) full simulation tools such as \geant~\cite{Geant4}, or by faster, yet less accurate, parametrised simulation tools such as \delphes~\cite{Delphes}.
This is where machine learning comes into play, with the hope of bridging the gap between the speed of the fast simulators and the accuracy of the full ones.

The full simulation pipeline can be broken down into several subtasks.
The first step is the event generation via matrix element calculation which simulates the hard process of the collision.
This is typically done using MC event generators such as \pythia~\cite{Pythia} or \madgraph~\cite{MadGraph}, and the output of this step is a collection of partons.
Next, the partons are used to simulate lower energy QCD processes such as radiation and hadronisation.
This occurs before they hit the detector material and greatly increases the number of particles produced by the collision.
Following this, all stable particles are propagated through the detector material, where they can interact with it and produce secondary particles.
This showering process is typically the most computationally expensive part of the simulation.
Finally, the deposited energy in the detector is digitised and reconstructed into objects such as jets, leptons and missing transverse energy (MET).

Several approaches have been proposed to replace parts of the pipeline with machine learning models, starting from the matrix element calculation and hard scattering event generation~\cite{butter2019ganlhc, maitre2021factorisation, winterhalder2022multiloop, bishara2023mlamplitudes, heimel2023madnis, heimel2023madnisreloaded, badger2023loop}.
Other approaches focus on the time-consuming material interactions and detector response parts, both in terms of image~\cite{oliveira2017learningbyexample, atlas2024deepmodels, paganini2018calogan, krause2021caloflow, cresswell2022caloman, mikuni2022caloscore, kobylianskii2024calograph, liu2024calovq, favaro2024calodream} and point cloud~\cite{kansal2021mpgan, kach2022jetflow, buhmann2023epicgan, kach2023attention, schnake2024calopointflow2} generation, with score-based generative models being applied with particular success to the latter~\cite{pcjedi, pcdroid, epicly, mikuni2023fpcd}.
While these are valid and well-motivated uses of machine learning, one can argue that a more ambitious goal would be to replace most of the simulation chain from hadronisation, showering and detector simulation to digitization and reconstruction~\cite{bellagente2020invertible, howard2022otus, turbosim, turbo, soybelman2023setgeneration, butter2023precision, butter2023jetgpt}.
The model would generate the reconstructed objects from random noise and may include conditioning by the partons from the event generator.
We present a novel method called \pippinfull (\pippin) to directly generate variable length full event point clouds at detector-level from parton-level information.

\begin{figure}[t]
    \includegraphics[scale=\diagramscale,trim={0 0 2cm 0},clip]{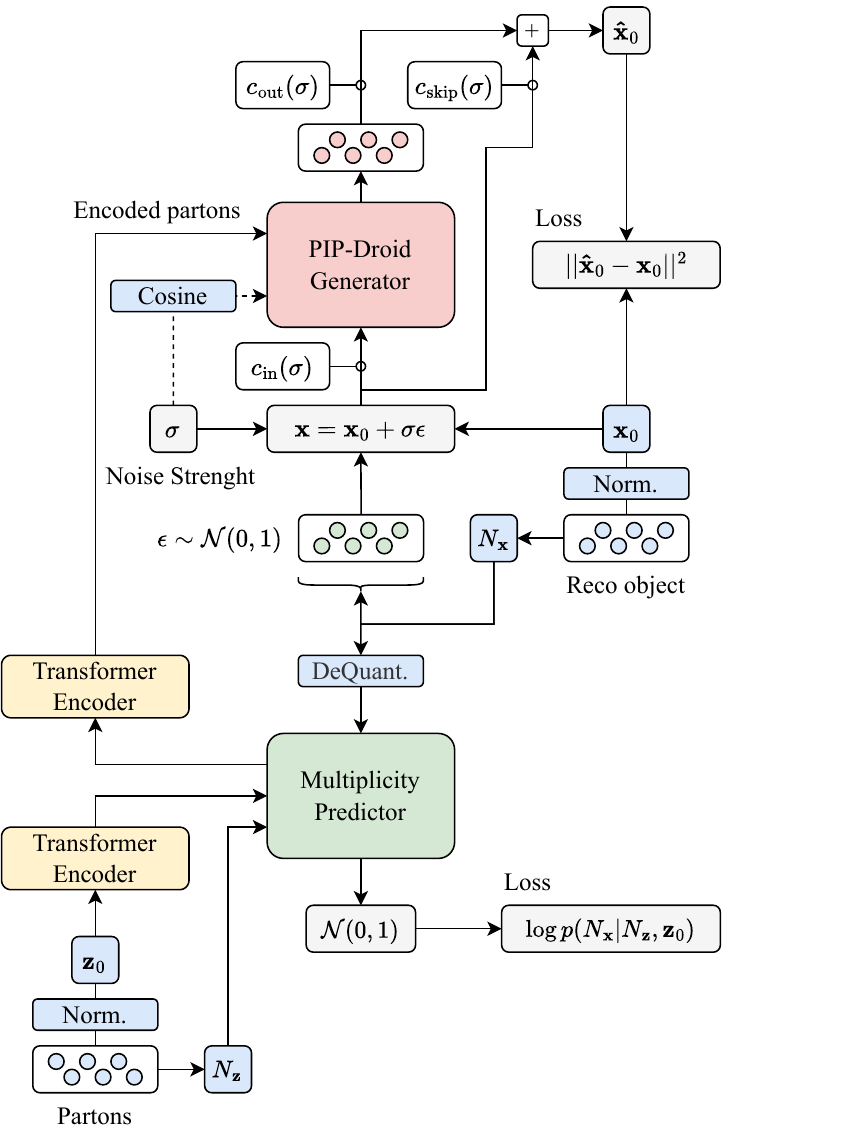}
    \caption{
        Diagram of the global architecture and the training processes of the \pippin model.
        It is made of two Transformer Encoders, which encode the partons, a Multiplicity Predictor, which predicts the number of reconstructed objects, and a PIP-Droid Generator, which conditionally generate these reconstructed objects.
    }
    \label{fig:pippin_training}
\end{figure}

The \pippin model is based on the combination of transformers~\cite{vaswani2023attention}, score-based models~\cite{karras2022elucidating}, and normalizing flows~\cite{durkan2019neural}.
A peculiarity of our model is its permutation invariance, an especially relevant feature to process unordered sets of particles.
In addition, such sets of particles can have a variable number of elements, which is a challenge that \pippin takes up by predicting the number of outputs it has to generate.
This means that we do not need to train a separate model for each multiplicity, truncate the output to a desired number of particles or capitalise on autoregressive generation.
On the contrary, the model can directly generate any number of particles while preserving correlations based on the input.
Furthermore, this property allows to generate events with a fixed number of particles of each type, which can be useful for filtering the simulation by restraining the phase space.
Another important feature of the \pippin model is its conditional nature, which allows either to change the properties of the generated events depending on the chosen input particles or to fix them in order to stochastically generate multiple instances from the same input.

A related approach can be found in~\cite{soybelman2023setgeneration}, with the main differences being that a) our inputs contain multiple partons of different natures instead of a single quark, b) our dataset is much larger but contains single pairs of input and output rather than several resimulated outputs per input, and c) our model uses score-based training and generation.

The case-study for our method is the simulation of top quark pair events in proton-proton collisions at the Large Hadron Collider (LHC), whose final state partons are used as input to the model to produce the full event at detector-level, also denoted as reconstructed objects.
The main contributions of this paper are:%
\footnote{Code is available at \url{https://github.com/rodem-hep/pippin}. \\
Data is available at \url{https://zenodo.org/records/12117432}.}
\begin{itemize}
    \item The introduction of a novel method to generate variable length full events at detector-level from parton-level information, leveraging cutting-edge transformers, score-based models and normalizing flows.
    \item A detailed study of the performance of the \pippin model in the context of top quark pair production at the LHC.
    \item The description and release of a new inclusive dataset of top quark pair events.
\end{itemize}
It should be noted that, despite the ambitions behind such an approach, a fundamental limitation arises whenever one wishes to generate full events rather than individual objects, namely the process-dependent nature of the simulation.
We leave studies on the exploitation of the generalisation capabilities of the \pippin model, through its inclusive extension to a wider range of processes, to future work.

\section{Dataset}
In this work, we focus on the simulation of pairs of top quarks~($t\bar{t}$) produced in proton-proton collisions at a centre of mass energy $\sqrt{s}=13$~TeV.
The hard interactions as well as parton shower and hadronisation is performed using \pythia~(v8.307)~\cite{Pythia} with the Monash tuned set of parameters~\cite{Monash}, using the \mbox{NNPDF2.3LO} PDF set~\cite{PartonDFs} in the LHAPDF~\cite{PartonDAccessLHCC} framework, at leading order accuracy in both QCD and Electroweak interactions.

During event generation no constraints are placed on the decays of the top quarks or $W$ bosons in the hard scatter, in order to capture the full range of final state particles.
Top quarks decay to a $W$ boson and a $b$-quark with a branching fraction over 99\%.
The $W$ bosons then decay to a pair of quarks or to a charged lepton and a neutrino.
Events can be categorised based on these decays into all-hadronic (0$\ell$), semi-leptonic ($1\ell$) and di-leptonic ($2\ell$) events.
In this categorisation we consider decays with tau leptons as semileptonic regardless of whether they decay hadronically or leptonically.

The detector response simulation is performed using Delphes~\cite{Delphes}~(v3.4.2) with a parametrisation consistent with the ATLAS detector~\cite{ATLAS}.

Jets are clustered using the anti-$k_t$ algorithm~\cite{AntiKt} with a radius parameter of $R=0.4$ using the FastJet package~\cite{FastJet}.
They are required to fall within $|\eta| < 2.5$ and to have a minimum transverse momentum $\pt>25$~GeV.
In addition to their four-momenta, jets are assigned a binary label corresponding to whether they pass a simulated $b$-jet identification algorithm, identifying jets which originate from $b$-hadrons ($b$-jets).
A similar $\tau$-jet identification algorithm is simulated, identifying jets which originate from hadronically decaying tau leptons.

Reconstructed electrons and muons are required to fall within $|\eta| < 2.5$ and to have a minimum transverse momentum $\pt>15$~GeV.
They are represented by their associated charge in addition to their four-momenta.
The reconstructed missing transverse momentum ($\vec{p}_\mathrm{T}^{\mathrm{miss}}$) of the event is calculated from the negative vector sum of all reconstructed visible particles.

All events with a minimum of 2 and a maximum of 16 jets are considered.
No requirement is placed on the number of $b$-jets or leptons, though only up to the leading two leptons ordered in descending \pt are considered.%
\footnote{This affects 0.002\% of all events, corresponding to 0.1\% of the events with at least two reconstructed leptons.}

In addition to the detector-level reconstructed objects, the four-momenta of the six parton-level final state particles (quarks, charged leptons and neutrinos) as well as for the two top quarks and $W$ bosons are kept for each event.
Truth association of reconstructed jets to the quarks from the $t\bar{t}$ decay (two quarks in the case of di-leptonic decays and up to six in the case of all-hadronic decays) is performed using $\Delta R$ matching.%
\footnote{From the pseudo-rapidity difference $\Delta\eta$ and the azimuthal angle difference $\Delta\phi$, one expresses the angular distance as $\Delta R = \sqrt{(\Delta\eta)^2 + (\Delta\phi)^2}$.}
Quarks are matched to jets within $\Delta R < 0.4$.
In order to remove ambiguity in the reconstruction of $W$ bosons and the top quarks from detector objects, an additional requirement is placed on events that no two jets are matched to the same parton, and no two partons are matched to the same jet.
It should be noted that this selection is not necessary for \pippin, and for a final model would be removed in order to preserve the event phase space from the original sample.%
\footnote{This removes 16.5\% of all events, with a higher fraction arising from the all-hadronic and semi-leptonic channels than di-leptonic, due to the higher jet and quark multiplicities.}

In total there are 50~million events, of which 40.4~million pass the selection criteria after reconstruction.
These correspond to approximately 18.6~million all-hadronic, 17.8~million semi-leptonic, and 4.1~million di-leptonic events based on the decays of the $t\bar{t}$ pair.
We use 37~million events for training and 0.8~million for validation of the \pippin model; the remaining 2.4~million events are reserved for evaluation.

\section{Method}
\label{sec:method}

\pippin is a conditional generative model that takes an unordered set of partons and outputs an unordered set of the reconstructed objects.
Each element of both sets is described by their four-momenta.
The model is made of four main components: two Transformer Encoders, the Multiplicity Predictor and the PIP-Droid Generator, as shown in \cref{fig:pippin_training}.

\begin{figure}[t]
    \includegraphics[scale=\diagramscale]{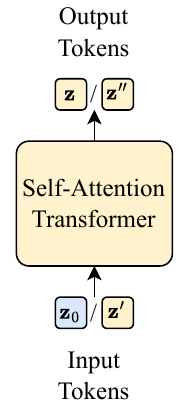}
    \includegraphics[scale=\diagramscale]{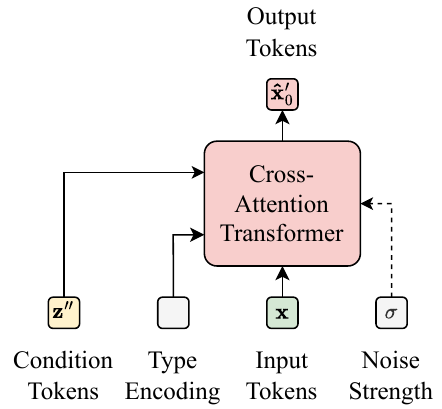}
    \caption{
        The Transformer Encoder architecture (left), which encodes the parton point cloud and the PIP-Droid Generator architecture (right), which denoises random tokens with type encoding into reconstructed objects, conditioned on the parton point cloud.
    }
    \label{fig:encoder_generator_light}
\end{figure}

The Transformer Encoders are two nearly identical\footnote{Only the input dimension differs.} self-attention transformers \cite{vaswani2023attention}, as shown in \cref{fig:encoder_generator_light}~(left), whose role is to encode the parton point cloud into latent representations further used by the other components of the model.
The representation from the first encoder is used as input to the Multiplicity Predictor, while the final representation from the second encoder is used as conditional input to the PIP-Droid Generator.

\begin{figure}[b]
    \includegraphics[scale=\diagramscale]{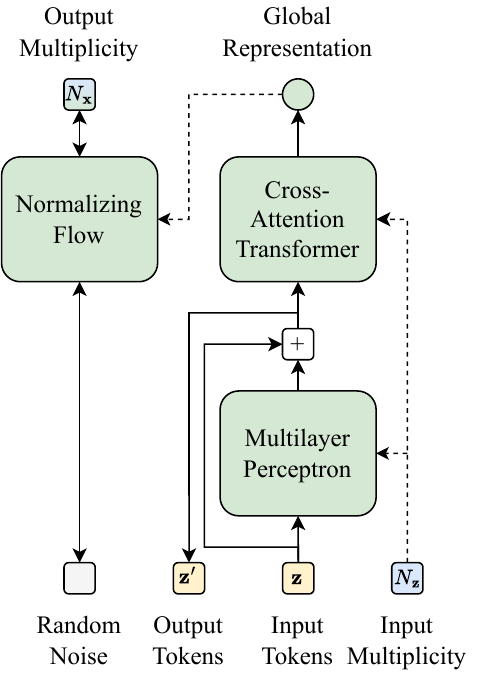}
    \caption{
        The Multiplicity Predictor architecture.
        The partons presence is predicted and residually added to the encoded parton point cloud.
        Then, a global representation of the new encoded parton point cloud is learnt and conditionally used to sample the output multiplicity.
    }
    \label{fig:multiplicity_predictor_light}
\end{figure}

The role of the Multiplicity Predictor is to estimate the number and type of reconstructed objects needed in the output point cloud.
It is composed of three blocks, as shown in \cref{fig:multiplicity_predictor_light}: a multilayer perceptron that predicts the probability that each of the input partons leads to a reconstructed object, a cross-attention transformer that extracts a global representation of the input point cloud, and a rational quadratic neural spline coupling normalizing flow \cite{durkan2019neural} that predicts the multiplicity of each particle type in the output point cloud.
Both the presence predictor and the global representation extractor are conditioned on the multiplicity of each particle type in the input point cloud, while the normalizing flow is conditioned on the extracted global representation.
We use the CDF-Dequantization method introduced in Ref.~\cite{schnake2024calopointflow2} in order to preprocess the discrete multiplicity into continuous data to train the normalizing flow.

Finally, the PIP-Droid Generator is an adapted version of the PC-Droid score-based model~\cite{pcjedi, pcdroid} that uses transformer decoder layers instead, as shown in \cref{fig:encoder_generator_light}~(right), in order to condition the denoising process using the encoded input point cloud.
As in Ref.~\cite{pcdroid}, we use the score-based framework of Ref.~\cite{karras2022elucidating} including how we precondition our network using $c_\text{in}$, $c_\text{out}$, and $c_\text{skip}$ functions in order to predict the original sample $x_0$ after perturbing it with scaled gaussian noise $x_t = x_0 + t \cdot \epsilon$, with $\epsilon \sim \mathcal{N}(0, I)$ and $t \in [0, 80]$.

As a conditional model, the PIP-Droid Generator takes as input the encoded parton point cloud, as well as the number of output tokens and their associated particle type, as given by the Multiplicity Predictor.
During training, the true multiplicity is used for the first two epochs as the Multiplicity Predictor training is delayed for stability purpose.
During generation, the predicted number of random tokens are sampled from a standard normal distribution and fed into the PIP-Droid Generator, which performs several denoising steps leading to the final output point cloud.
In addition, a learnable token, specific to each particle type, is added to the input at each step as we observed that without this, the training was prone to collapse.

More details on the internal architecture and the training parameters can be found in \cref{app:architecture_details}.
Several important features of the \pippin model are worth highlighting here.

Due to the permutation equivariant nature of transformers and the permutation invariant cross-attention operation used by the Multiplicity Predictor and the PIP-Droid Generator, the model as a whole is invariant to permutations of the input point cloud.
This property is crucial when dealing with sets of particles, such as a couple of leptons and several jets, for which any imposed order is weakly motivated.
In addition, since it is trained on multiple top quark decay channels at once, the model must handle this wide variety of inputs and outputs.
The multiplicity distribution depends on the decay channel, which further increases the variability of the output point clouds.
By learning to generate all these different types of point cloud, the \pippin model is expected to have a great capacity for generalisation.

One advantage of conditional generation is that the input point cloud can be modified before it is fed into the model, allowing the output point cloud to be generated accordingly.
One can add partons to the input, or change their energy or direction, and generate new events based on these changes.
Moreover, by removing the Multiplicity Predictor from the generation pipeline, one can use the model to generate point clouds with a fixed number of particles of each type, filtering the simulation and allowing for increased control over the generation process.
The Multiplicity Predictor also predicts the probability for a parton to be present in the detector-level reconstructed point cloud.
Partons can be lost by falling outside the acceptance region of the detector, or due to reconstruction inefficiency, because the parton is not energetic enough to be distinguished from noise, or is too close to another parton to be resolved.
This additional task greatly assists the model in learning the correct number of particles to generate, as well as their kinematic properties, since partons provide a strong target for the reconstructed objects.
The additional reconstructed objects are still conditionally generated, since they initially originate from parton radiations, but the model can learn that they are not the direct deposit of a parton.

The attention mechanism at the heart of transformers allows each token to attend to every other token, meaning that the values of the output tokens are all correlated.
This is relevant to generate unordered sets of particles, as it allows the model to learn correlations between them and thus to generate more realistic point clouds.
In addition, compared to an autoregressive method, the \pippin model does not create a hierarchy between the particles, which is better suited to the unordered nature of the data.
Moreover, since both the Multiplicity Predictor and the PIP-Droid Generator are conditional networks that take as input random noise, the model can be used to generate multiple outputs from the same input point cloud.
Two main advantages come from this property.
First, it may allow the model the properly learn the distinction between the content of the data, namely the physical nature of the point clouds such as the type of particles and features, and the style of the data, namely the statistical properties of the point clouds such as the distribution of the features.
Second, it can be used to estimate part of the systematic uncertainty of the generation, by generating multiple instances of the same process and assessing its variability.

\section{Results}
The results are presented following the model pipeline.
First, we present the estimations of the Multiplicity Predictor both in terms of parton presence in the reconstructed objects and the multiplicity of the outputs.
Second, we show the outputs of the PIP-Droid Generator, i.e. the reconstructed object kinematics.
Third, we perform an additional study of the kinematics of the underlying particles that initiated the partons.
Note that for each plot in this section, the uncertainties for \pippin are estimated as the standard deviation over 5 generated sets initialised with different seeds, while for MC they are assumed to follow a Poisson distribution.

\subsection{Multiplicity prediction}

\begin{figure}[t]
    \includegraphics[scale=\plotscale]{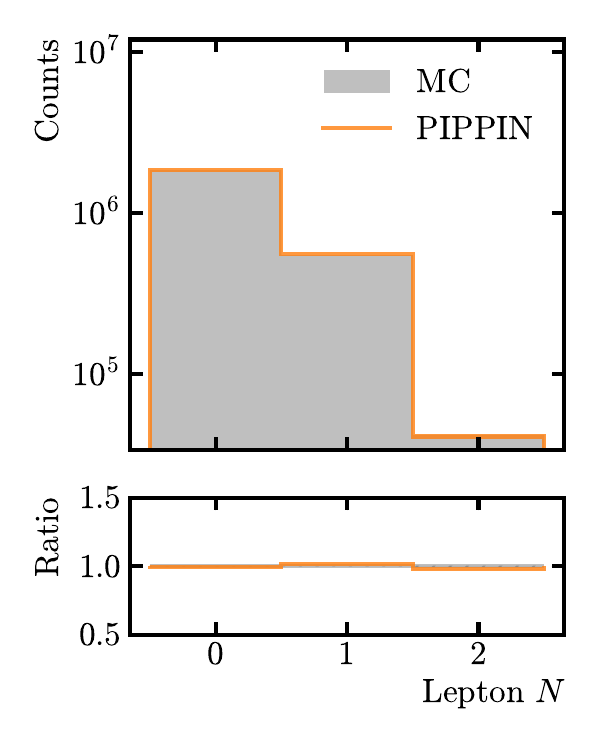}
    \includegraphics[scale=\plotscale]{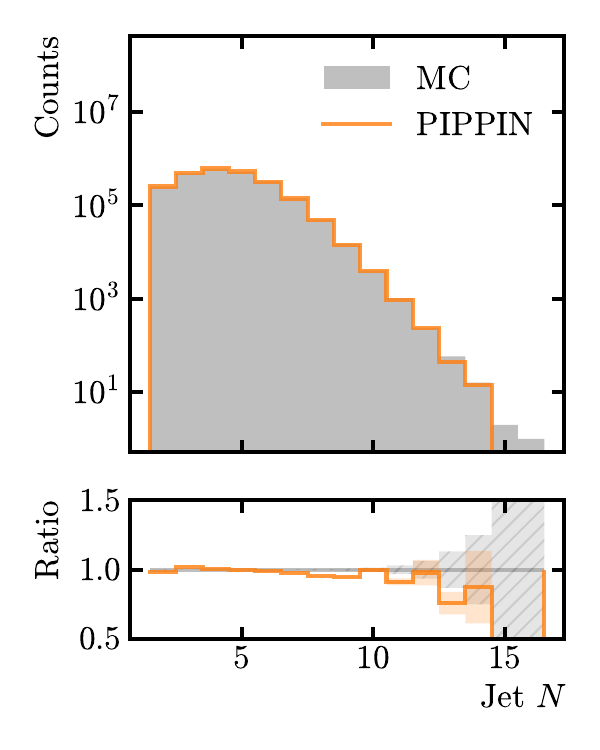}
    \caption{
        Marginal distributions of the learnt reco-level multiplicities.
        \textbf{Left:} The multiplicity of the leptons in the reconstructed objects.
        \textbf{Right:} The multiplicity of the jets in the reconstructed objects.
        The grey area corresponds to the original MC simulation and the orange line to the output of the \pippin model.
        The bottom plots show the ratios of the histograms with respect to MC and the uncertainties as shaded areas.
    }
    \label{fig:marginals_n}
\end{figure}

\begin{figure}[b]
    \includegraphics[scale=\plotscale]{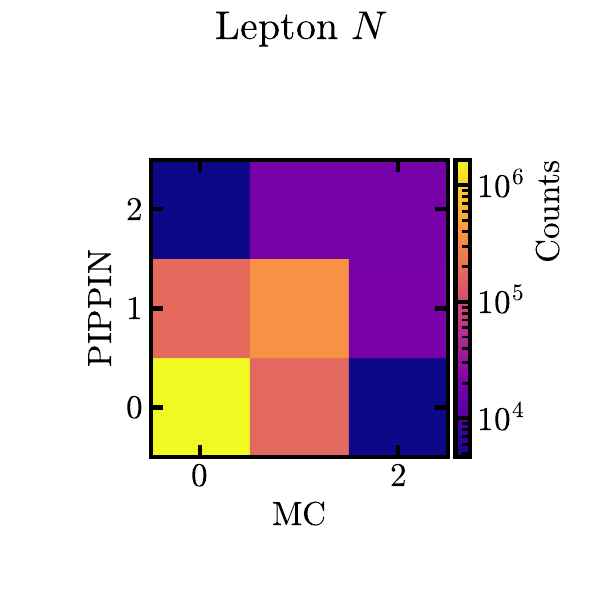}
    \includegraphics[scale=\plotscale]{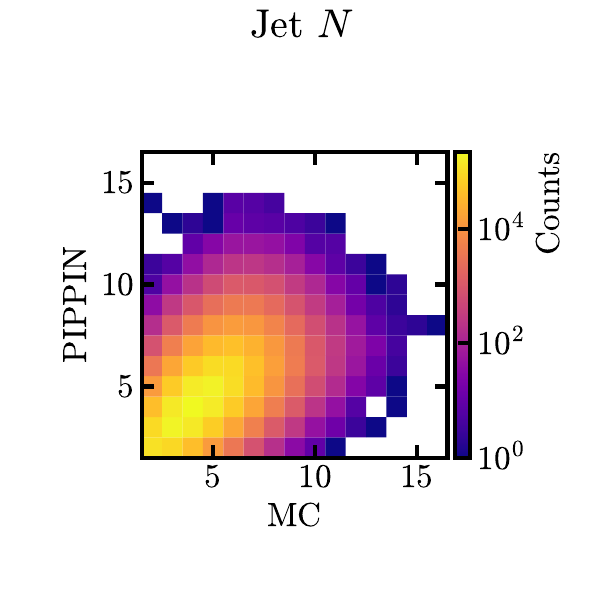}
    \caption{
        2D marginal distributions of the learnt reco-level multiplicities.
        \textbf{Left:} The multiplicity of the leptons in the reconstructed objects.
        \textbf{Right:} The multiplicity of the jets in the reconstructed objects.
        The $x$-axis corresponds to the original MC simulation and the $y$-axis to the associated output of the \pippin model.
    }
    \label{fig:marginals_2D_n}
\end{figure}

The first main step of the \pippin model is to predict the number of particles of each type that should be present in the reconstructed point cloud based on the input point cloud.
\cref{fig:marginals_n} shows the marginal distributions of the predicted multiplicity $N$ for the leptons and the jets.
One can observe a perfect agreement between the predicted and true lepton multiplicity.
The jet multiplicity is also nearly perfectly predicted overall, but the model tends to slightly underestimate their number, having difficulties with events containing more than 14 jets.
However, this tail of the distribution represents a tiny fraction of the events.

The corresponding 2D marginal distributions shown in \cref{fig:marginals_2D_n} give more insight into how the input partons influence the predicted reconstructed object multiplicity.
We recall here that the Multiplicity Predictor is conditioned on the input parton multiplicity and an encoded representation of the partons themselves, together with the estimated probability of presence of each of them in the output.
A perfect agreement between the truth and the prediction would be a diagonal.
We observe that indeed most events are close to it.
However, due to the intrinsic stochasticity of the process, a spread around the diagonal is inevitable, as clearly seen for the jet multiplicity in \cref{fig:marginals_2D_n}~(right).
The model is therefore not expected to predict the exact truth, but rather to learn the distribution of the reconstructed object multiplicity given the input partons.
The 2D histograms confirm this expected behaviour.

\begin{figure}[t]
    \includegraphics[scale=\plotscale]{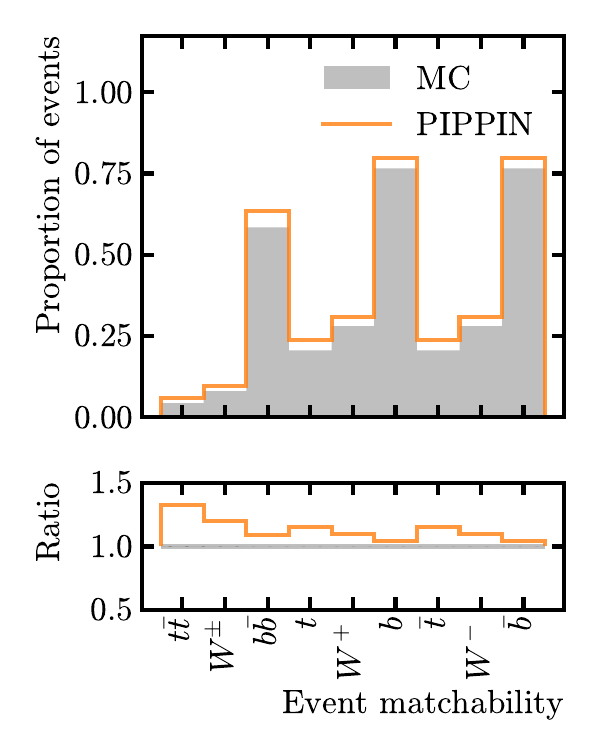}
    \includegraphics[scale=\plotscale]{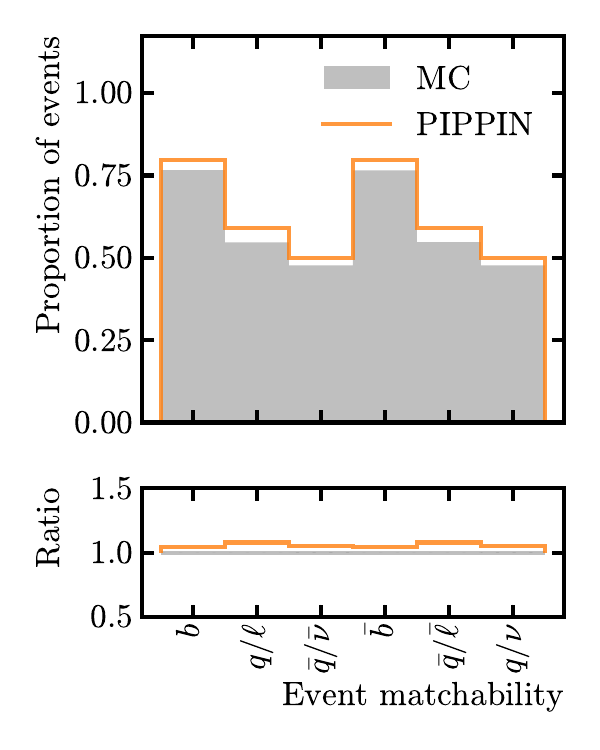}
    \caption{
        Marginal distributions of the learnt event-level matchability.
        \textbf{Left:} The matchability for the intermediate underlying particles and the $b$-quarks.
        \textbf{Right:} The matchability for the final state particles, namely the partons.
        The grey area corresponds to the original MC simulation and the orange line to the output of the \pippin model.
        The bottom plots show the ratios of the histograms with respect to MC and the uncertainties as shaded, but invisible, areas.
    }
    \label{fig:marginals_match}
\end{figure}

In addition, \cref{fig:marginals_match} shows a selected summary of the presence prediction, or matchability, gathered by underlying intermediate and final particles.
A particle is considered present, or matched, if all its decay products, or partons, are present in the reconstructed objects of the event.
As mentioned above, a parton may be missing due to acceptance and efficiency effects, namely detector blind regions, noise thresholds or resolution limitations.
On the left plot, from left to right, the first bin shows the proportion of events for which all 6 partons are matched to form the $t\bar{t}$ pair, the second bin for which the 4 partons originating from the $W$ bosons are matched, the third bin for which the 2 $b$-quarks are matched, etc.
Intuitively, the $b$ and $\bar{b}$-quark bins are expected to be the most populated, as they require a single parton to be matched to a reconstructed object per event.
The $b\bar{b}$ bin contains the intersection of the two and is therefore expected to be slightly less populated.
The $W$ boson bins also require two partons to be matched, one of which being sometimes a neutrino.
This adds a layer of complexity to the matching, which is reflected in the lower population of these bins.
The $t$ and $\bar{t}$ bins being the intersection of the $W$ boson and $b$-quark bins, they are also slightly less populated.
Finally, the $t\bar{t}$ bin requires all 6 partons to be matched and is therefore the least populated.
On the right plot, the bins simply show the proportion of events for which each of the 6 partons are individually matched.

We observe a reasonably good agreement between the predicted and true presence, with a slight overestimation in all cases.
This means that the model has a conservative tendency, trying to keep the partons in the reconstructed objects more often than the imperfect detector response would allow.
The exact reasons behind this behaviour are not clear, but one has to keep in mind that the histograms show the matchability as a binary variable, while the model predicts a probability of presence for each parton.
Furthermore, the presence prediction is not a critical quantity for the model to produce accurate output point clouds, but rather a helpful auxiliary task.
Therefore, the observed discrepancy is not a major concern.

\subsection{Kinematic properties}

\begin{figure}[t]
    \includegraphics[scale=\plotscale]{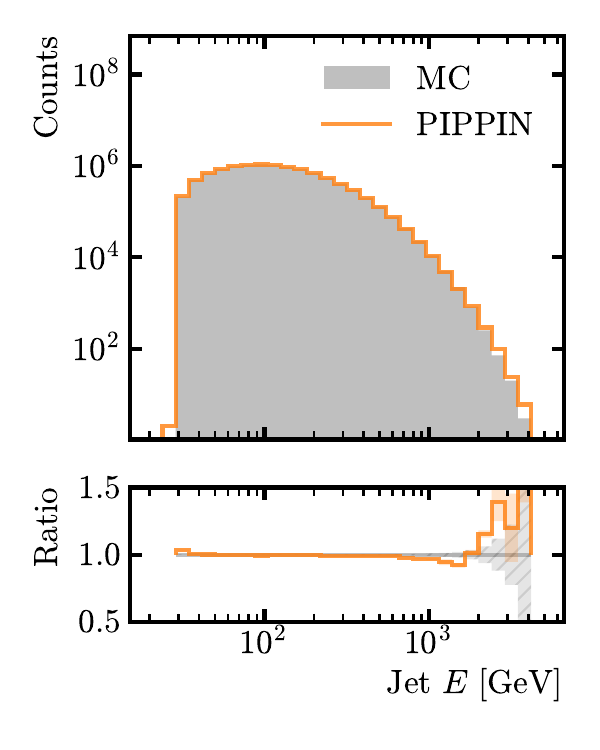}
    \includegraphics[scale=\plotscale]{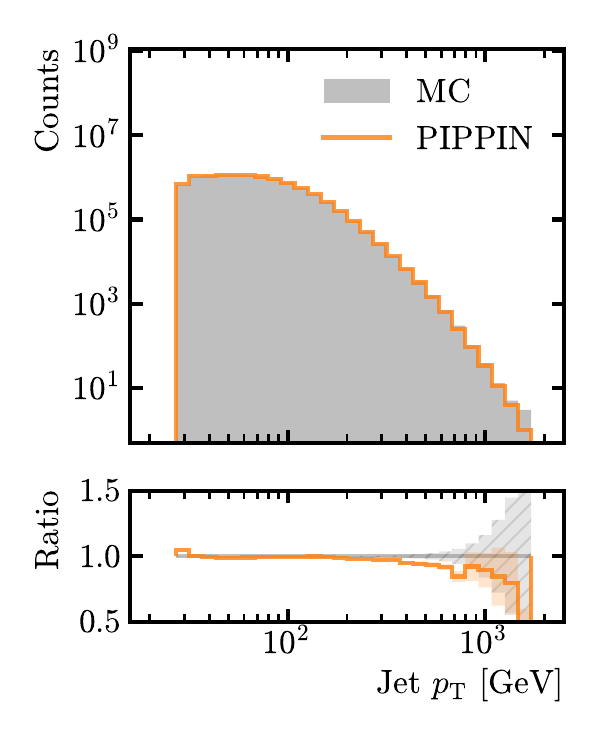}
    \caption{
        Marginal distributions of the learnt features of the reco-level jets.
        \textbf{Left:} The energy of the jets in the reconstructed objects.
        \textbf{Right:} The $p_\mathrm{T}$ of the jets in the reconstructed objects.
        The grey area corresponds to the original MC simulation and the orange line to the output of the \pippin model.
        The bottom plots show the ratios of the histograms with respect to MC and the uncertainties as shaded areas.
    }
    \label{fig:marginals_jet}
\end{figure}

Each particle in the point clouds is represented by a set of four features: its transverse momentum, pseudo-rapidity, azimuthal angle, and mass (for the partons) or energy (for the reconstructed objects).
The marginal distributions of the energy and transverse momentum are shown in \cref{fig:marginals_jet} for the jets and in \cref{app:additional_plots} for the other reconstructed objects.
Apart from an underestimated number of high-energy and transverse momentum jets, and an overestimate of very high-energy jets, likely due to a low statistics in the tails of the distributions, and an overestimate of very low-energy and transverse momentum jets, likely due to the sharp cut at zero, one can observe a very good agreement between the predicted and true histograms.
This good agreement is expected as the PIP-Droid Generator score-based model is trained on these features.

\begin{figure}[b]
    \includegraphics[scale=\plotscale]{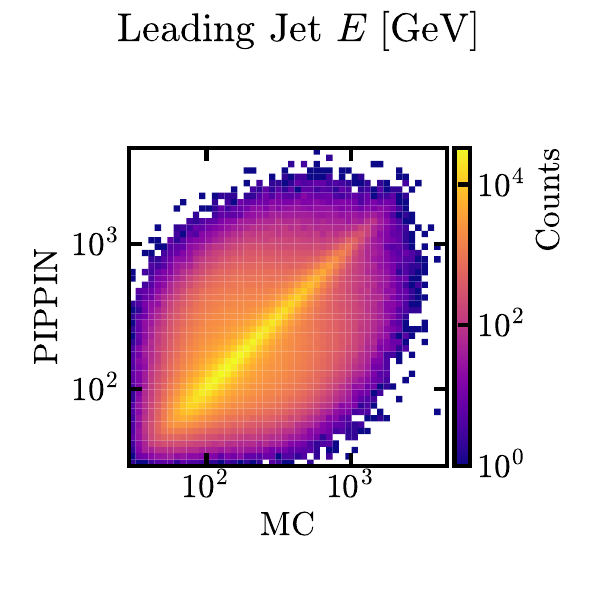}
    \includegraphics[scale=\plotscale]{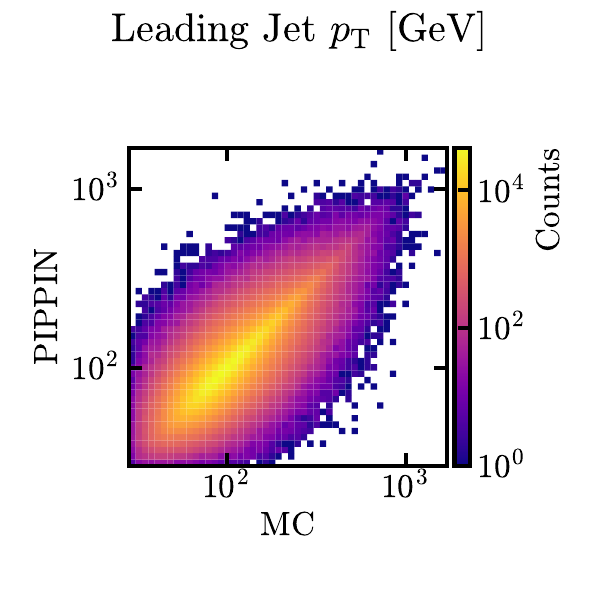}
    \caption{
        2D marginal distributions of the learnt features of the reco-level jets.
        \textbf{Left:} The energy of the jets in the reconstructed objects.
        \textbf{Right:} The $p_\mathrm{T}$ of the jets in the reconstructed objects.
        The $x$-axis corresponds to the original MC simulation and the $y$-axis to the associated output of the \pippin model.
    }
    \label{fig:marginals_2D_jet}
\end{figure}

The 2D marginal distributions in \cref{fig:marginals_2D_jet} lead to the same conclusion.
The diagonals are well populated, with an expected spread due to the stochastic nature of the process.

\subsection{Underlying particles}
\label{sec:results_underlying}

\begin{figure*}[t]
    \includegraphics[scale=\plotscale]{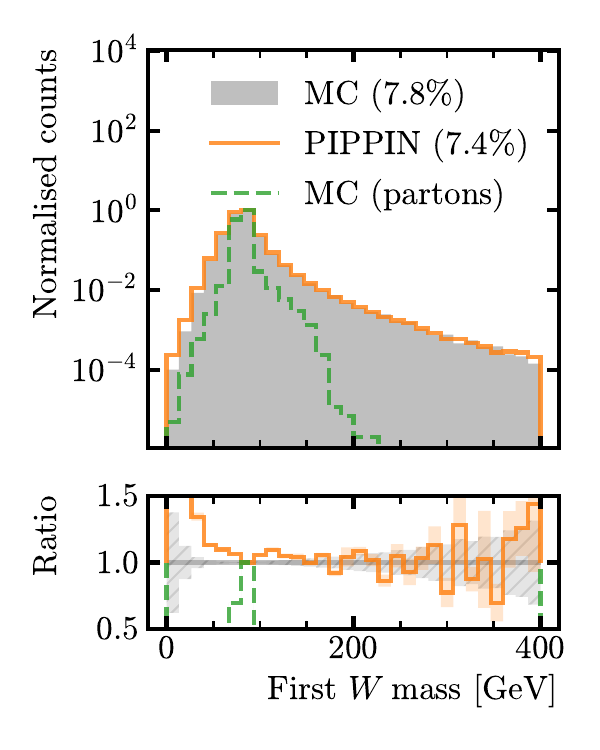}
    \includegraphics[scale=\plotscale]{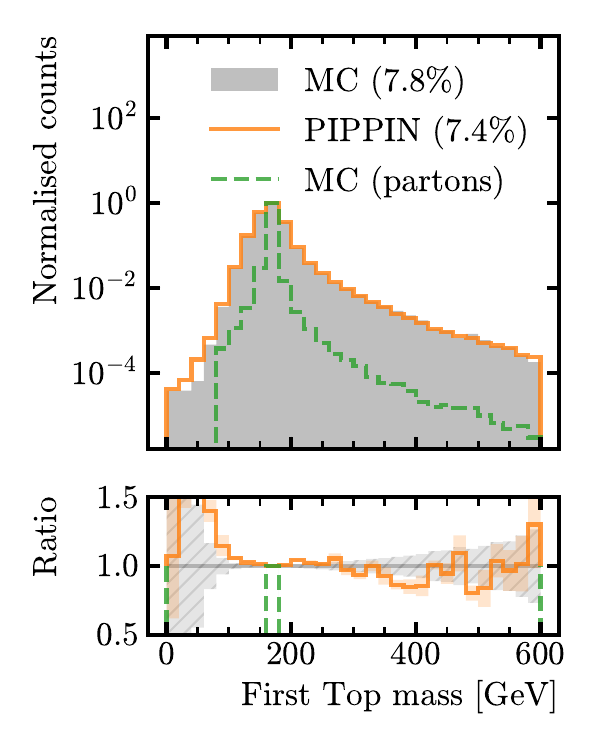}
    \includegraphics[scale=\plotscale]{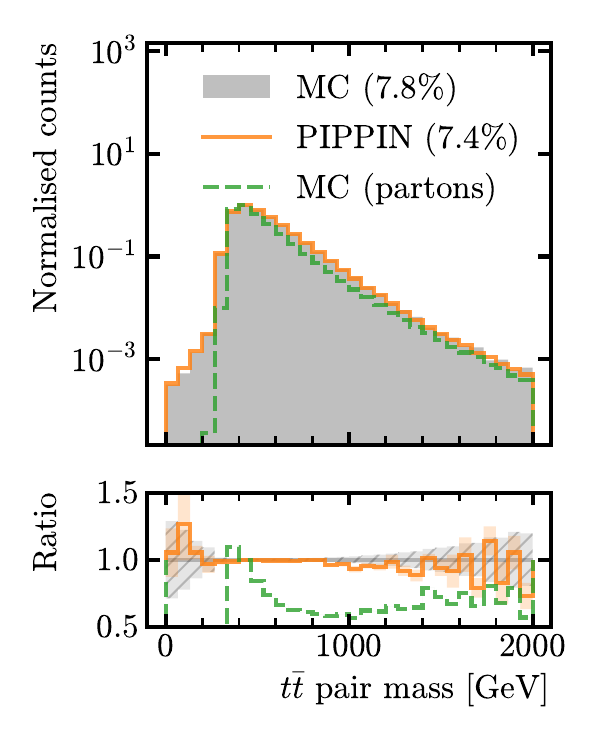}
    \caption{
        Marginal distributions of the invariant masses of the underlying particles at reco-level.
        \textbf{Left:} The mass of the first reconstructed $W$ boson.
        \textbf{Middle:} The mass of the first reconstructed top quark.
        \textbf{Right:} The mass of the whole reconstructed $t\bar{t}$ system.
        By first we mean the particle originating from the top quark, as opposed to the anti-top quark.
        The grey area corresponds to the original MC reco-level simulation, the orange line to the output of the \pippin model and the dashed green line to the MC parton-level simulation on which the model is conditioned.
        The bottom plots show the ratios of the histograms with respect to MC and the uncertainties as shaded areas.
        The percentages indicate the proportion of events for which all partons are unambiguously matched and therefore present on the plots.
    }
    \label{fig:masses}
\end{figure*}

\begin{figure*}[t]
    \includegraphics[scale=\plotscale]{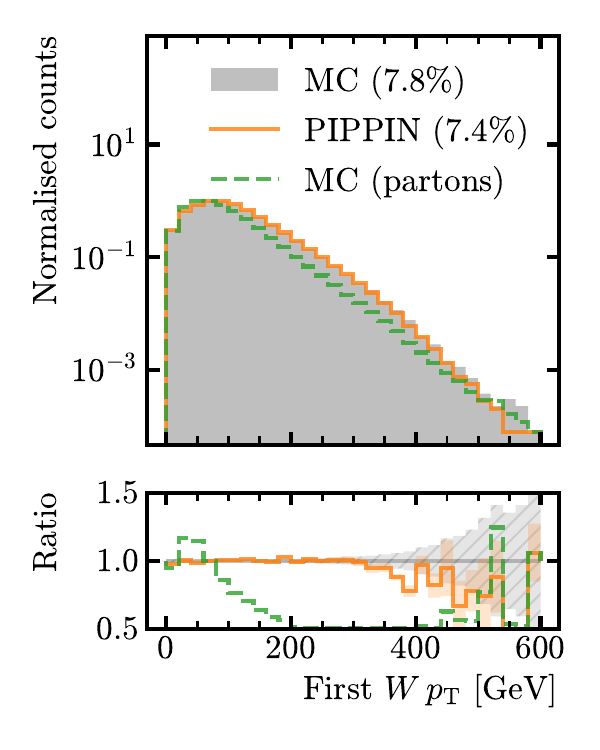}
    \includegraphics[scale=\plotscale]{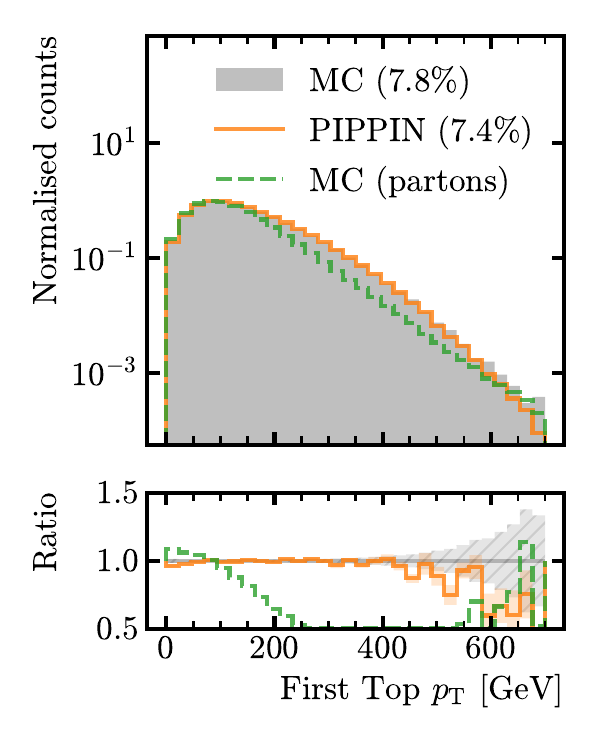}
    \includegraphics[scale=\plotscale]{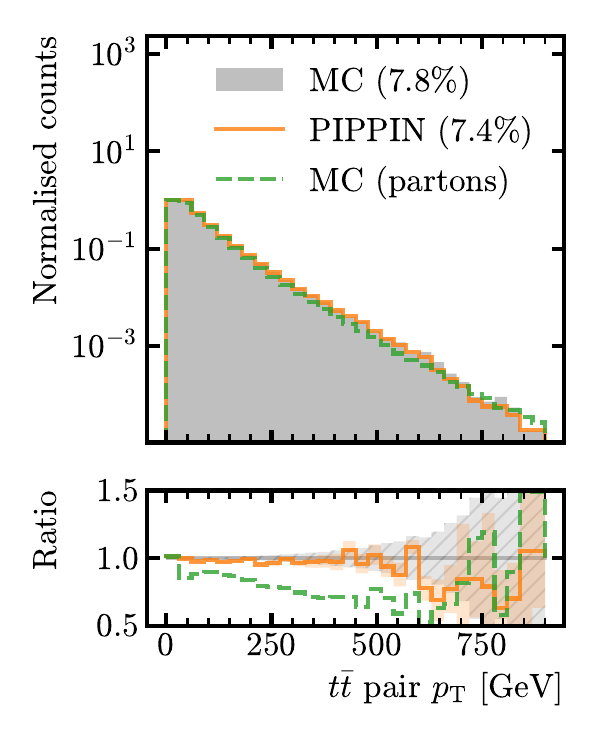}
    \caption{
        Marginal distributions of the transverse momenta of the underlying particles at reco-level.
        \textbf{Left:} The transverse momentum of the first reconstructed $W$ boson.
        \textbf{Middle:} The transverse momentum of the first reconstructed top quark.
        \textbf{Right:} The transverse momentum of the whole reconstructed $t\bar{t}$ system.
        By first we mean the particle originating from the top quark, as opposed to the anti-top quark.
        The grey area corresponds to the original MC reco-level simulation, the orange line to the output of the \pippin model and the dashed green line to the MC parton-level simulation on which the model is conditioned.
        The bottom plots show the ratios of the histograms with respect to MC and the uncertainties as shaded areas.
        The percentages indicate the proportion of events for which all partons are unambiguously matched and therefore present on the plots.
    }
    \label{fig:momenta}
\end{figure*}

In addition to the features of the reconstructed objects, we can compute several properties of the underlying decayed particles, namely the invariant masses and transverse momenta of the $W$ bosons, the top and anti-top quarks, and the whole $t\bar{t}$ system.
In order to identify the objects originating from each underlying particle, we perform a matching between the partons and the reconstructed objects based on their proximity in the $(\eta, \phi)$ plane.
Similarly to how the dataset is produced, objects with $\Delta R < 0.4$ are matched to partons, keeping only fully matched events and discarding those with multiple objects matched to a parton, and vice versa.
Furthermore, di-leptonic events originally contain two neutrinos, which have to be disentangled from the single MET object present at the detector level.
To avoid the need of any complex algorithm, we create two fake detector-level neutrinos by taking the highest $p_\mathrm{T}$ parton-level neutrino and the remaining MET after subtraction.
For both the semi-leptonic and di-leptonic cases, we also use the parton-level neutrino system to extract the pseudo-rapidity associated to the MET, and then compute the corresponding energy as $E = p_\mathrm{T} \cdot \cosh \eta$.
The use of these approximate proxy variables is not a problem as we are only interested in the \pippin capabilities to reproduce them and not in their precise estimation.

The marginal distributions of the invariant masses and transverse momenta of the underlying particles are shown in \cref{fig:masses,fig:momenta}, where we also overlay the parton-level distributions for comparison purpose.
In essence, the \pippin model aims to transform the parton-level distributions into the reco-level ones, conditioned on these partons.
The percentage of fully matched events, i.e., events for which we can recover all the underlying particles, is shown in the legend of each plot.
It is expected to correspond to the height of the $t\bar{t}$ bin in \cref{fig:marginals_match} (left), as this represents the proportion of events for which the 6 partons are predicted to be present in the reconstructed objects.
Note that the percentages are small, indicating that more than 92\% of the time at least one parton is missing due to the aforementioned detector acceptance and efficiency limitations.
It is interesting to notice that the truth events percentage is almost reached by the \pippin generated events.
This is a good indicator that the model is able to generate different event topologies in the same proportion as the truth.
The apparent contradiction between the lower percentage of fully matched events generated by \pippin shown in \cref{fig:masses,fig:momenta} and the higher proportion of events for which all partons are predicted to be present in \cref{fig:marginals_match} (left) with respect to MC should not be a concern.
On the contrary, this is a good sign that the PIP-Droid Generator is able to compensate for a possible mismodelling of the presence predicted by the Multiplicity Predictor, justifying our claim that this is not a critical quantity to predict.

Although there is an tendency to overestimate the number of low mass particles at the expense of high mass ones, the distributions of the invariant masses in \cref{fig:masses} and the transverse momenta in \cref{fig:momenta} show a good agreement between the truth and the prediction.
It is worth highlighting this agreement with the detector-level distributions, bearing in mind that the model is conditioned on the partons whose corresponding distributions are significantly different.
This is very encouraging, as these quantities are not used as training targets, while their modelling is crucial for the detector-level point cloud to be physically meaningful.
Additional plots of invariant masses and transverse momenta can be found in \cref{app:additional_plots}.

\subsection{Comparison to other models}

\pippin is the first model applied to the problem of simulating an inclusive set of top quark pair production events of variable length at the LHC.
This means that for now the only way to compare our method to other models is to restrict the set of events to the specific topology considered by them.
We can compare to the OTUS~\cite{howard2022otus} and Turbo-Sim~\cite{turbosim} related models, which are both designed to simulate the semi-leptonic decay of a top quark pair with exactly 1 lepton, MET and 4 jets present in the reconstructed objects.
Plots comparing the features of the dataset originally used by these two models and the restricted version of the dataset used in this paper are provided in \cref{app:additional_plots}.

\begin{table}[t]
    \begin{tabular}{llllllll}
        \toprule
        & \multicolumn{3}{c}{Reco. objects} & \multicolumn{4}{c}{Underlying particles} \\
        \cmidrule(l){2-4} \cmidrule(l){5-8}
        \textbf{Model} & \boldmath{$p_y^{jet_1}$} & \boldmath{$p_z^{jet_1}$} & \boldmath{$E^{jet_1}$} & \boldmath{$m^{t\bar{t}}$} & \boldmath{$m^{W_1}$} & \boldmath{$m^{t_1}$} & \boldmath{$m^{t_2}$} \\
        \midrule
        OTUS            & \it{3.78} & \it{2.39} & \it{5.75} & \it{15.8} & \it{11.7} & \it{14.1} & \it{24.9} \\
        Turbo-Sim       & \it{2.89} & \it{10.3} & \it{4.43} & \it{2.97} & \it{7.72} & \it{5.20} & \it{8.52} \\
        Turbo-Sim (new) & 8.63 & 12.6 & 6.32 & 7.90* & 38.8* & 38.5* & 43.6* \\
        \pippin         & \bf{0.32} & \bf{0.33} & \bf{0.34} & \bf{4.00} & \bf{3.66} & \bf{3.27} & \bf{2.44} \\
        \pippin (inc)   & \it{0.08} & \it{0.14} & \it{0.12} & \it{0.33} & \it{1.69} & \it{0.54} & \it{0.60} \\
        \bottomrule
    \end{tabular}
    \caption{
        The Kolmogorov-Smirnov distances [$\times 10^{-2}$] computed between the original MC reco-level simulation and the output of different models.
        Lower values mean better agreement.
        The observables considered are the momentum in the $y$ and $z$ directions and the energy of the leading reconstructed jet in $p_\mathrm{T}$, as well as the mass of several underlying particles.
        For OTUS and Turbo-Sim, $m^{W_1}$, $m^{t_1}$ and $m^{t_2}$ are the masses of the hadronic $W$ boson, the leptonic top quark and the hadronic top quark, respectively.
        For \pippin, they are the masses of the first $W$ boson, the first top quark, and the second top quark, respectively, as defined in \cref{fig:masses,fig:momenta}.
        The mass of the whole $t\bar{t}$ system is denoted $m^{t\bar{t}}$ for all models.
        The OTUS and Turbo-Sim lines show the original results of these models based on the dataset provided by the authors.
        The Turbo-Sim (new) and \pippin lines show the results of these models on the restricted version of the dataset presented in this study.
        The \pippin (inc) line shows the results for \pippin when not restricted to semi-leptonic events containing exactly 1 lepton, MET and 4 jets.
        We have marked with an asterisk results where the matching procedure was slightly relaxed in order to capture enough events for comparison.
        The matching radius was increased to $\Delta R < 0.8$ and events with multiple matching reconstructed objects per parton were allowed, with the closest being selected.
    }
    \label{tab:ks_comparison}
\end{table}

\begin{figure}[t]
    \includegraphics[scale=\plotscale]{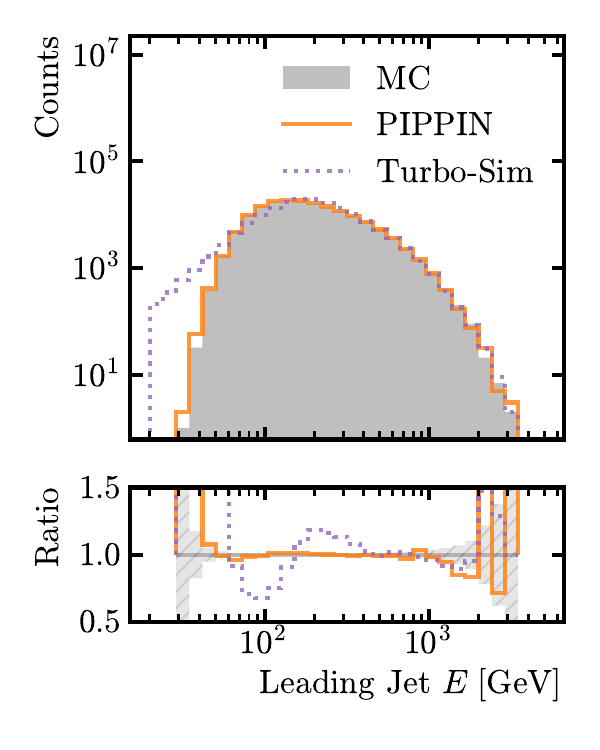}
    \includegraphics[scale=\plotscale]{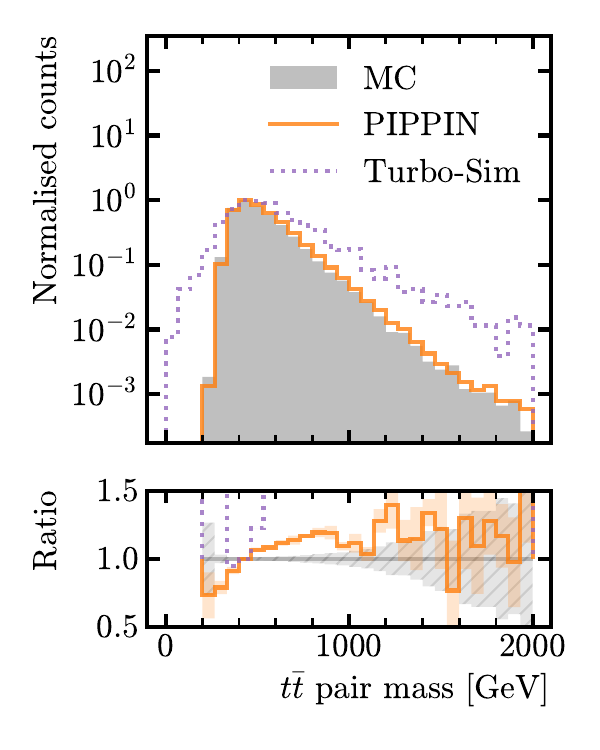}
    \caption{
        Comparison of the marginal distributions for different models.
        \textbf{Left:} The energy of the leading jet in the reconstructed objects.
        \textbf{Right:} The mass of the whole reconstructed $t\bar{t}$ system.
        The grey area corresponds to the original MC simulation, the orange line to the output of the \pippin model and the purple dotted line to the output of the Turbo-Sim model.
        The bottom plots show the ratios of the histograms with respect to MC and the uncertainties as shaded areas.
    }
    \label{fig:comparison}
\end{figure}

In \cref{tab:ks_comparison}, we show the Kolmogorov-Smirnov (KS) distances, which reflect the agreement between two cumulative distribution functions, between the original MC reco-level simulation and the output of the different models.
The observables considered are divided in two categories, namely the kinematic features of the reconstructed point cloud and the features of the underlying intermediate particles.
For the point cloud, we show the KS distances for the $p_\mathrm{T}$ leading jet momentum in the $y$ and $z$ directions as well as for its energy.
For the underlying particles, we show the KS distances for the mass of the $t\bar{t}$ system, one of the $W$ bosons, and the two top quarks.
\cref{fig:comparison} shows a comparison of the distributions of the leading jet energy and the $t\bar{t}$ mass for the MC simulation as well as the output of \pippin and Turbo-Sim.
We show both the results when the output events produced by \pippin are selected to match the topology of the other models, and when all events are considered regardless of topology.
We also show the original results of the OTUS and Turbo-Sim models based on the dataset provided by the authors, as well as the results of the Turbo-Sim model when retrained on the restricted version of the dataset presented in this study.
Note that the Turbo-Sim model was retrained using the same architecture and hyperparameters as the original model, without further optimisation for the new dataset.
The worse results compared to the original Turbo-Sim are therefore to be expected.
We observe that \pippin significantly outperforms the two other models.
It is also worth noting that, without the topological restriction, the results are even better, which has two meanings.
On the one hand, it may indicate that the subset of events considered by the other models is more difficult to simulate.
This would not be surprising as the majority of events in the dataset are all-hadronic and do not contain charged leptons and neutrinos, removing an important source of complexity.
On the other hand, it means that \pippin is able to accurately simulate a much wider range of events than the other models, while still outperforming them on the specific subset.
It would be interesting to see how \pippin compares to itself when trained directly on the restricted dataset, rather than trained inclusively and then restricted on the outputs, but this goes beyond the scope of this work and is left for future studies.

\section{Conclusion}
In this work, we have presented a new approach to the simulation of high-energy physics events, leveraging cutting-edge deep learning architectures and a shift of paradigm, namely the direct generation of reconstructed objects from final state partons instead of the traditional step-by-step decomposition of the pipeline.
We have shown that the proposed \pippin model is able to handle the complexity of the top quark pair production problem and to correctly predict the multiplicity and kinematics of the outputs.
It is worth emphasising that the model is able to learn the topologies of several decay channels simultaneously, without the need for channel-specific training.
Moreover, the correlations between the different simulated reconstructed objects, tested by exploring the kinematics of the underlying decayed particles, are well captured by the model.
This is a crucial feature for the accurate simulation of high-energy physics events, as the kinematics of the final state particles are not independent of each other.
The control over the generated events is also a key feature of the model.
It could allow to fix some properties of the outputs, as the number of jets and leptons or the presence of specific partons in the reconstructed objects.

It would be interesting to make use of this controllability to further investigate the generalisation capabilities of the model, by testing it on modified or unseen processes.
For example, one could train the model on a given Beyond Standard Model process and test it on the same process parametrised with different masses or couplings.
One could even test the model on a completely different process, such as the production of a new particle alongside the top quark pairs, and see how well it generalises.

Lastly, the method presented in this work is nicely suited for the reverse problem, namely the unfolding of the reconstructed objects to the initial partons.
If several final states are allowed, it would be a natural extension of the \pippin model, as the tasks involved would be analogous.
The model would have to predict the multiplicity of each parton and the correspondence between the reconstructed objects and the partons.
Then, the conditional generation would follow the same procedure, with the partons as outputs and the reconstructed objects as conditional inputs.
An analogous task, which however stops the unfolding before reversing the hadronisation and radiation processes, has been studied in~\cite{shmakov2024VLD} during the development of this work with a similar pipeline that uses a variational latent diffusion generator instead.
We leave the complete study of applying our method to the unfolding problem for future work.

\section*{Acknowledgements}
The authors would like to acknowledge funding through the SNSF Sinergia grant CRSII5\_193716 ``Robust Deep Density Models for High-Energy Particle Physics and Solar Flare Analysis (RODEM)'', the SNSF project grant 200020\_212127 ``At the two upgrade frontiers: machine learning and the ITk Pixel detector'' and the ESKAS Scholarship for international students.

The computations were performed at the University of Geneva on the ``Baobab'' and ``Yggdrasil'' HPC clusters.

\clearpage
\bibliography{bib/biblio.bib}

\begin{thebibliography}{46}%
\makeatletter
\providecommand \@ifxundefined [1]{%
 \@ifx{#1\undefined}
}%
\providecommand \@ifnum [1]{%
 \ifnum #1\expandafter \@firstoftwo
 \else \expandafter \@secondoftwo
 \fi
}%
\providecommand \@ifx [1]{%
 \ifx #1\expandafter \@firstoftwo
 \else \expandafter \@secondoftwo
 \fi
}%
\providecommand \natexlab [1]{#1}%
\providecommand \enquote  [1]{``#1''}%
\providecommand \bibnamefont  [1]{#1}%
\providecommand \bibfnamefont [1]{#1}%
\providecommand \citenamefont [1]{#1}%
\providecommand \href@noop [0]{\@secondoftwo}%
\providecommand \href [0]{\begingroup \@sanitize@url \@href}%
\providecommand \@href[1]{\@@startlink{#1}\@@href}%
\providecommand \@@href[1]{\endgroup#1\@@endlink}%
\providecommand \@sanitize@url [0]{\catcode `\\12\catcode `\$12\catcode `\&12\catcode `\#12\catcode `\^12\catcode `\_12\catcode `\%12\relax}%
\providecommand \@@startlink[1]{}%
\providecommand \@@endlink[0]{}%
\providecommand \url  [0]{\begingroup\@sanitize@url \@url }%
\providecommand \@url [1]{\endgroup\@href {#1}{\urlprefix }}%
\providecommand \urlprefix  [0]{URL }%
\providecommand \Eprint [0]{\href }%
\providecommand \doibase [0]{https://doi.org/}%
\providecommand \selectlanguage [0]{\@gobble}%
\providecommand \bibinfo  [0]{\@secondoftwo}%
\providecommand \bibfield  [0]{\@secondoftwo}%
\providecommand \translation [1]{[#1]}%
\providecommand \BibitemOpen [0]{}%
\providecommand \bibitemStop [0]{}%
\providecommand \bibitemNoStop [0]{.\EOS\space}%
\providecommand \EOS [0]{\spacefactor3000\relax}%
\providecommand \BibitemShut  [1]{\csname bibitem#1\endcsname}%
\let\auto@bib@innerbib\@empty
\bibitem [{\citenamefont {Agostinelli}\ \emph {et~al.}(2003)\citenamefont {Agostinelli} \emph {et~al.}}]{Geant4}%
  \BibitemOpen
  \bibfield  {author} {\bibinfo {author} {\bibfnamefont {S.}~\bibnamefont {Agostinelli}} \emph {et~al.},\ }\bibfield  {title} {\bibinfo {title} {{GEANT4—a simulation toolkit}},\ }\href {https://doi.org/10.1016/S0168-9002(03)01368-8} {\bibfield  {journal} {\bibinfo  {journal} {Nuclear Instruments and Methods in Physics Research Section A: Accelerators, Spectrometers, Detectors and Associated Equipment}\ }\textbf {\bibinfo {volume} {506}},\ \bibinfo {pages} {250} (\bibinfo {year} {2003})}\BibitemShut {NoStop}%
\bibitem [{\citenamefont {{The DELPHES 3 collaboration}}\ \emph {et~al.}(2014)\citenamefont {{The DELPHES 3 collaboration}}, \citenamefont {de~Favereau}, \citenamefont {Delaere}, \citenamefont {Demin}, \citenamefont {Giammanco}, \citenamefont {Lemaître}, \citenamefont {Mertens},\ and\ \citenamefont {Selvaggi}}]{Delphes}%
  \BibitemOpen
  \bibfield  {author} {\bibinfo {author} {\bibnamefont {{The DELPHES 3 collaboration}}}, \bibinfo {author} {\bibfnamefont {J.}~\bibnamefont {de~Favereau}}, \bibinfo {author} {\bibfnamefont {C.}~\bibnamefont {Delaere}}, \bibinfo {author} {\bibfnamefont {P.}~\bibnamefont {Demin}}, \bibinfo {author} {\bibfnamefont {A.}~\bibnamefont {Giammanco}}, \bibinfo {author} {\bibfnamefont {V.}~\bibnamefont {Lemaître}}, \bibinfo {author} {\bibfnamefont {A.}~\bibnamefont {Mertens}},\ and\ \bibinfo {author} {\bibfnamefont {M.}~\bibnamefont {Selvaggi}},\ }\bibfield  {title} {\bibinfo {title} {{DELPHES 3: a modular framework for fast simulation of a generic collider experiment}},\ }\href {https://doi.org/10.1007/JHEP02(2014)057} {\bibfield  {journal} {\bibinfo  {journal} {Journal of High Energy Physics}\ }\textbf {\bibinfo {volume} {2014}},\ \bibinfo {pages} {57} (\bibinfo {year} {2014})}\BibitemShut {NoStop}%
\bibitem [{\citenamefont {Sjöstrand}\ \emph {et~al.}(2008)\citenamefont {Sjöstrand}, \citenamefont {Mrenna},\ and\ \citenamefont {Skands}}]{Pythia}%
  \BibitemOpen
  \bibfield  {author} {\bibinfo {author} {\bibfnamefont {T.}~\bibnamefont {Sjöstrand}}, \bibinfo {author} {\bibfnamefont {S.}~\bibnamefont {Mrenna}},\ and\ \bibinfo {author} {\bibfnamefont {P.}~\bibnamefont {Skands}},\ }\bibfield  {title} {\bibinfo {title} {{A brief introduction to PYTHIA 8.1}},\ }\href {https://doi.org/10.1016/j.cpc.2008.01.036} {\bibfield  {journal} {\bibinfo  {journal} {Computer Physics Communications}\ }\textbf {\bibinfo {volume} {178}},\ \bibinfo {pages} {852} (\bibinfo {year} {2008})}\BibitemShut {NoStop}%
\bibitem [{\citenamefont {Alwall}\ \emph {et~al.}(2014)\citenamefont {Alwall}, \citenamefont {Frederix}, \citenamefont {Frixione}, \citenamefont {Hirschi}, \citenamefont {Maltoni}, \citenamefont {Mattelaer}, \citenamefont {Shao}, \citenamefont {Stelzer}, \citenamefont {Torrielli},\ and\ \citenamefont {Zaro}}]{MadGraph}%
  \BibitemOpen
  \bibfield  {author} {\bibinfo {author} {\bibfnamefont {J.}~\bibnamefont {Alwall}}, \bibinfo {author} {\bibfnamefont {R.}~\bibnamefont {Frederix}}, \bibinfo {author} {\bibfnamefont {S.}~\bibnamefont {Frixione}}, \bibinfo {author} {\bibfnamefont {V.}~\bibnamefont {Hirschi}}, \bibinfo {author} {\bibfnamefont {F.}~\bibnamefont {Maltoni}}, \bibinfo {author} {\bibfnamefont {O.}~\bibnamefont {Mattelaer}}, \bibinfo {author} {\bibfnamefont {H.-S.}\ \bibnamefont {Shao}}, \bibinfo {author} {\bibfnamefont {T.}~\bibnamefont {Stelzer}}, \bibinfo {author} {\bibfnamefont {P.}~\bibnamefont {Torrielli}},\ and\ \bibinfo {author} {\bibfnamefont {M.}~\bibnamefont {Zaro}},\ }\bibfield  {title} {\bibinfo {title} {{The automated computation of tree-level and next-to-leading order differential cross sections, and their matching to parton shower simulations}},\ }\href {https://doi.org/10.1007/JHEP07(2014)079} {\bibfield  {journal} {\bibinfo  {journal} {Journal of High Energy Physics}\ }\textbf {\bibinfo {volume} {2014}},\ \bibinfo {pages} {79} (\bibinfo {year} {2014})}\BibitemShut {NoStop}%
\bibitem [{\citenamefont {Butter}\ \emph {et~al.}(2019)\citenamefont {Butter}, \citenamefont {Plehn},\ and\ \citenamefont {Winterhalder}}]{butter2019ganlhc}%
  \BibitemOpen
  \bibfield  {author} {\bibinfo {author} {\bibfnamefont {A.}~\bibnamefont {Butter}}, \bibinfo {author} {\bibfnamefont {T.}~\bibnamefont {Plehn}},\ and\ \bibinfo {author} {\bibfnamefont {R.}~\bibnamefont {Winterhalder}},\ }\bibfield  {title} {\bibinfo {title} {{How to GAN LHC events}},\ }\href {https://doi.org/10.21468/SciPostPhys.7.6.075} {\bibfield  {journal} {\bibinfo  {journal} {SciPost Physics}\ }\textbf {\bibinfo {volume} {7}},\ \bibinfo {pages} {075} (\bibinfo {year} {2019})}\BibitemShut {NoStop}%
\bibitem [{\citenamefont {Maître}\ and\ \citenamefont {Truong}(2021)}]{maitre2021factorisation}%
  \BibitemOpen
  \bibfield  {author} {\bibinfo {author} {\bibfnamefont {D.}~\bibnamefont {Maître}}\ and\ \bibinfo {author} {\bibfnamefont {H.}~\bibnamefont {Truong}},\ }\bibfield  {title} {\bibinfo {title} {{A factorisation-aware Matrix element emulator}},\ }\href {https://doi.org/10.1007/JHEP11(2021)066} {\bibfield  {journal} {\bibinfo  {journal} {Journal of High Energy Physics}\ }\textbf {\bibinfo {volume} {2021}},\ \bibinfo {pages} {66} (\bibinfo {year} {2021})}\BibitemShut {NoStop}%
\bibitem [{\citenamefont {Winterhalder}\ \emph {et~al.}(2022)\citenamefont {Winterhalder}, \citenamefont {Magerya}, \citenamefont {Villa}, \citenamefont {Jones}, \citenamefont {Kerner}, \citenamefont {Butter}, \citenamefont {Heinrich},\ and\ \citenamefont {Plehn}}]{winterhalder2022multiloop}%
  \BibitemOpen
  \bibfield  {author} {\bibinfo {author} {\bibfnamefont {R.}~\bibnamefont {Winterhalder}}, \bibinfo {author} {\bibfnamefont {V.}~\bibnamefont {Magerya}}, \bibinfo {author} {\bibfnamefont {E.}~\bibnamefont {Villa}}, \bibinfo {author} {\bibfnamefont {S.}~\bibnamefont {Jones}}, \bibinfo {author} {\bibfnamefont {M.}~\bibnamefont {Kerner}}, \bibinfo {author} {\bibfnamefont {A.}~\bibnamefont {Butter}}, \bibinfo {author} {\bibfnamefont {G.}~\bibnamefont {Heinrich}},\ and\ \bibinfo {author} {\bibfnamefont {T.}~\bibnamefont {Plehn}},\ }\bibfield  {title} {\bibinfo {title} {{Targeting multi-loop integrals with neural networks}},\ }\href {https://doi.org/10.21468/SciPostPhys.12.4.129} {\bibfield  {journal} {\bibinfo  {journal} {SciPost Physics}\ }\textbf {\bibinfo {volume} {12}},\ \bibinfo {pages} {129} (\bibinfo {year} {2022})}\BibitemShut {NoStop}%
\bibitem [{\citenamefont {Bishara}\ and\ \citenamefont {Montull}(2023)}]{bishara2023mlamplitudes}%
  \BibitemOpen
  \bibfield  {author} {\bibinfo {author} {\bibfnamefont {F.}~\bibnamefont {Bishara}}\ and\ \bibinfo {author} {\bibfnamefont {M.}~\bibnamefont {Montull}},\ }\bibfield  {title} {\bibinfo {title} {{Machine learning amplitudes for faster event generation}},\ }\href {https://doi.org/10.1103/PhysRevD.107.L071901} {\bibfield  {journal} {\bibinfo  {journal} {Physical Review D}\ }\textbf {\bibinfo {volume} {107}},\ \bibinfo {pages} {L071901} (\bibinfo {year} {2023})}\BibitemShut {NoStop}%
\bibitem [{\citenamefont {Heimel}\ \emph {et~al.}(2023{\natexlab{a}})\citenamefont {Heimel}, \citenamefont {Winterhalder}, \citenamefont {Butter}, \citenamefont {Isaacson}, \citenamefont {Krause}, \citenamefont {Maltoni}, \citenamefont {Mattelaer},\ and\ \citenamefont {Plehn}}]{heimel2023madnis}%
  \BibitemOpen
  \bibfield  {author} {\bibinfo {author} {\bibfnamefont {T.}~\bibnamefont {Heimel}}, \bibinfo {author} {\bibfnamefont {R.}~\bibnamefont {Winterhalder}}, \bibinfo {author} {\bibfnamefont {A.}~\bibnamefont {Butter}}, \bibinfo {author} {\bibfnamefont {J.}~\bibnamefont {Isaacson}}, \bibinfo {author} {\bibfnamefont {C.}~\bibnamefont {Krause}}, \bibinfo {author} {\bibfnamefont {F.}~\bibnamefont {Maltoni}}, \bibinfo {author} {\bibfnamefont {O.}~\bibnamefont {Mattelaer}},\ and\ \bibinfo {author} {\bibfnamefont {T.}~\bibnamefont {Plehn}},\ }\bibfield  {title} {\bibinfo {title} {{MadNIS - Neural multi-channel importance sampling}},\ }\href {https://doi.org/10.21468/SciPostPhys.15.4.141} {\bibfield  {journal} {\bibinfo  {journal} {SciPost Physics}\ }\textbf {\bibinfo {volume} {15}},\ \bibinfo {pages} {141} (\bibinfo {year} {2023}{\natexlab{a}})}\BibitemShut {NoStop}%
\bibitem [{\citenamefont {Heimel}\ \emph {et~al.}(2023{\natexlab{b}})\citenamefont {Heimel}, \citenamefont {Huetsch}, \citenamefont {Maltoni}, \citenamefont {Mattelaer}, \citenamefont {Plehn},\ and\ \citenamefont {Winterhalder}}]{heimel2023madnisreloaded}%
  \BibitemOpen
  \bibfield  {author} {\bibinfo {author} {\bibfnamefont {T.}~\bibnamefont {Heimel}}, \bibinfo {author} {\bibfnamefont {N.}~\bibnamefont {Huetsch}}, \bibinfo {author} {\bibfnamefont {F.}~\bibnamefont {Maltoni}}, \bibinfo {author} {\bibfnamefont {O.}~\bibnamefont {Mattelaer}}, \bibinfo {author} {\bibfnamefont {T.}~\bibnamefont {Plehn}},\ and\ \bibinfo {author} {\bibfnamefont {R.}~\bibnamefont {Winterhalder}},\ }\href@noop {} {\bibinfo {title} {{The MadNIS Reloaded}}} (\bibinfo {year} {2023}{\natexlab{b}}),\ \Eprint {https://arxiv.org/abs/2311.01548} {arXiv:2311.01548 [hep-ph]} \BibitemShut {NoStop}%
\bibitem [{\citenamefont {Badger}\ \emph {et~al.}(2023)\citenamefont {Badger}, \citenamefont {Butter}, \citenamefont {Luchmann}, \citenamefont {Pitz},\ and\ \citenamefont {Plehn}}]{badger2023loop}%
  \BibitemOpen
  \bibfield  {author} {\bibinfo {author} {\bibfnamefont {S.}~\bibnamefont {Badger}}, \bibinfo {author} {\bibfnamefont {A.}~\bibnamefont {Butter}}, \bibinfo {author} {\bibfnamefont {M.}~\bibnamefont {Luchmann}}, \bibinfo {author} {\bibfnamefont {S.}~\bibnamefont {Pitz}},\ and\ \bibinfo {author} {\bibfnamefont {T.}~\bibnamefont {Plehn}},\ }\bibfield  {title} {\bibinfo {title} {{Loop amplitudes from precision networks}},\ }\href {https://doi.org/10.21468/SciPostPhysCore.6.2.034} {\bibfield  {journal} {\bibinfo  {journal} {SciPost Physics Core}\ }\textbf {\bibinfo {volume} {6}},\ \bibinfo {pages} {034} (\bibinfo {year} {2023})}\BibitemShut {NoStop}%
\bibitem [{\citenamefont {de~Oliveira}\ \emph {et~al.}(2017)\citenamefont {de~Oliveira}, \citenamefont {Paganini},\ and\ \citenamefont {Nachman}}]{oliveira2017learningbyexample}%
  \BibitemOpen
  \bibfield  {author} {\bibinfo {author} {\bibfnamefont {L.}~\bibnamefont {de~Oliveira}}, \bibinfo {author} {\bibfnamefont {M.}~\bibnamefont {Paganini}},\ and\ \bibinfo {author} {\bibfnamefont {B.}~\bibnamefont {Nachman}},\ }\bibfield  {title} {\bibinfo {title} {{Learning Particle Physics by Example: Location-Aware Generative Adversarial Networks for Physics Synthesis}},\ }\href {https://doi.org/10.1007/s41781-017-0004-6} {\bibfield  {journal} {\bibinfo  {journal} {Computing and Software for Big Science}\ }\textbf {\bibinfo {volume} {1}},\ \bibinfo {pages} {4} (\bibinfo {year} {2017})}\BibitemShut {NoStop}%
\bibitem [{\citenamefont {{The ATLAS Collaboration}}(2024)}]{atlas2024deepmodels}%
  \BibitemOpen
  \bibfield  {author} {\bibinfo {author} {\bibnamefont {{The ATLAS Collaboration}}},\ }\bibfield  {title} {\bibinfo {title} {{Deep Generative Models for Fast Photon Shower Simulation in ATLAS}},\ }\href {https://doi.org/10.1007/s41781-023-00106-9} {\bibfield  {journal} {\bibinfo  {journal} {Computing and Software for Big Science}\ }\textbf {\bibinfo {volume} {8}},\ \bibinfo {pages} {7} (\bibinfo {year} {2024})}\BibitemShut {NoStop}%
\bibitem [{\citenamefont {Paganini}\ \emph {et~al.}(2018)\citenamefont {Paganini}, \citenamefont {de~Oliveira},\ and\ \citenamefont {Nachman}}]{paganini2018calogan}%
  \BibitemOpen
  \bibfield  {author} {\bibinfo {author} {\bibfnamefont {M.}~\bibnamefont {Paganini}}, \bibinfo {author} {\bibfnamefont {L.}~\bibnamefont {de~Oliveira}},\ and\ \bibinfo {author} {\bibfnamefont {B.}~\bibnamefont {Nachman}},\ }\bibfield  {title} {\bibinfo {title} {{CaloGAN: Simulating 3D high energy particle showers in multilayer electromagnetic calorimeters with generative adversarial networks}},\ }\href {https://doi.org/10.1103/PhysRevD.97.014021} {\bibfield  {journal} {\bibinfo  {journal} {Physical Review D}\ }\textbf {\bibinfo {volume} {97}},\ \bibinfo {pages} {014021} (\bibinfo {year} {2018})}\BibitemShut {NoStop}%
\bibitem [{\citenamefont {Krause}\ and\ \citenamefont {Shih}(2021)}]{krause2021caloflow}%
  \BibitemOpen
  \bibfield  {author} {\bibinfo {author} {\bibfnamefont {C.}~\bibnamefont {Krause}}\ and\ \bibinfo {author} {\bibfnamefont {D.}~\bibnamefont {Shih}},\ }\href@noop {} {\bibinfo {title} {{CaloFlow II: Even Faster and Still Accurate Generation of Calorimeter Showers with Normalizing Flows}}} (\bibinfo {year} {2021}),\ \Eprint {https://arxiv.org/abs/2110.11377} {arXiv:2110.11377 [physics.ins-det]} \BibitemShut {NoStop}%
\bibitem [{\citenamefont {Cresswell}\ \emph {et~al.}(2022)\citenamefont {Cresswell}, \citenamefont {Ross}, \citenamefont {Loaiza-Ganem}, \citenamefont {Reyes-Gonzalez}, \citenamefont {Letizia},\ and\ \citenamefont {Caterini}}]{cresswell2022caloman}%
  \BibitemOpen
  \bibfield  {author} {\bibinfo {author} {\bibfnamefont {J.~C.}\ \bibnamefont {Cresswell}}, \bibinfo {author} {\bibfnamefont {B.~L.}\ \bibnamefont {Ross}}, \bibinfo {author} {\bibfnamefont {G.}~\bibnamefont {Loaiza-Ganem}}, \bibinfo {author} {\bibfnamefont {H.}~\bibnamefont {Reyes-Gonzalez}}, \bibinfo {author} {\bibfnamefont {M.}~\bibnamefont {Letizia}},\ and\ \bibinfo {author} {\bibfnamefont {A.~L.}\ \bibnamefont {Caterini}},\ }\href@noop {} {\bibinfo {title} {{CaloMan: Fast generation of calorimeter showers with density estimation on learned manifolds}}} (\bibinfo {year} {2022}),\ \Eprint {https://arxiv.org/abs/2211.15380} {arXiv:2211.15380 [hep-ph]} \BibitemShut {NoStop}%
\bibitem [{\citenamefont {Mikuni}\ and\ \citenamefont {Nachman}(2022)}]{mikuni2022caloscore}%
  \BibitemOpen
  \bibfield  {author} {\bibinfo {author} {\bibfnamefont {V.}~\bibnamefont {Mikuni}}\ and\ \bibinfo {author} {\bibfnamefont {B.}~\bibnamefont {Nachman}},\ }\bibfield  {title} {\bibinfo {title} {{Score-based generative models for calorimeter shower simulation}},\ }\href {https://doi.org/10.1103/PhysRevD.106.092009} {\bibfield  {journal} {\bibinfo  {journal} {Physical Review D}\ }\textbf {\bibinfo {volume} {106}},\ \bibinfo {pages} {092009} (\bibinfo {year} {2022})}\BibitemShut {NoStop}%
\bibitem [{\citenamefont {Kobylianskii}\ \emph {et~al.}(2024)\citenamefont {Kobylianskii}, \citenamefont {Soybelman}, \citenamefont {Dreyer},\ and\ \citenamefont {Gross}}]{kobylianskii2024calograph}%
  \BibitemOpen
  \bibfield  {author} {\bibinfo {author} {\bibfnamefont {D.}~\bibnamefont {Kobylianskii}}, \bibinfo {author} {\bibfnamefont {N.}~\bibnamefont {Soybelman}}, \bibinfo {author} {\bibfnamefont {E.}~\bibnamefont {Dreyer}},\ and\ \bibinfo {author} {\bibfnamefont {E.}~\bibnamefont {Gross}},\ }\href@noop {} {\bibinfo {title} {{CaloGraph: Graph-based diffusion model for fast shower generation in calorimeters with irregular geometry}}} (\bibinfo {year} {2024}),\ \Eprint {https://arxiv.org/abs/2402.11575} {arXiv:2402.11575 [hep-ex]} \BibitemShut {NoStop}%
\bibitem [{\citenamefont {Liu}\ \emph {et~al.}(2024)\citenamefont {Liu}, \citenamefont {Shimmin}, \citenamefont {Liu}, \citenamefont {Shlizerman}, \citenamefont {Li},\ and\ \citenamefont {Hsu}}]{liu2024calovq}%
  \BibitemOpen
  \bibfield  {author} {\bibinfo {author} {\bibfnamefont {Q.}~\bibnamefont {Liu}}, \bibinfo {author} {\bibfnamefont {C.}~\bibnamefont {Shimmin}}, \bibinfo {author} {\bibfnamefont {X.}~\bibnamefont {Liu}}, \bibinfo {author} {\bibfnamefont {E.}~\bibnamefont {Shlizerman}}, \bibinfo {author} {\bibfnamefont {S.}~\bibnamefont {Li}},\ and\ \bibinfo {author} {\bibfnamefont {S.-C.}\ \bibnamefont {Hsu}},\ }\href@noop {} {\bibinfo {title} {{Calo-VQ: Vector-Quantized Two-Stage Generative Model in Calorimeter Simulation}}} (\bibinfo {year} {2024}),\ \Eprint {https://arxiv.org/abs/2405.06605} {arXiv:2405.06605 [physics.ins-det]} \BibitemShut {NoStop}%
\bibitem [{\citenamefont {Favaro}\ \emph {et~al.}(2024)\citenamefont {Favaro}, \citenamefont {Ore}, \citenamefont {Schweitzer},\ and\ \citenamefont {Plehn}}]{favaro2024calodream}%
  \BibitemOpen
  \bibfield  {author} {\bibinfo {author} {\bibfnamefont {L.}~\bibnamefont {Favaro}}, \bibinfo {author} {\bibfnamefont {A.}~\bibnamefont {Ore}}, \bibinfo {author} {\bibfnamefont {S.~P.}\ \bibnamefont {Schweitzer}},\ and\ \bibinfo {author} {\bibfnamefont {T.}~\bibnamefont {Plehn}},\ }\href@noop {} {\bibinfo {title} {{CaloDREAM -- Detector Response Emulation via Attentive flow Matching}}} (\bibinfo {year} {2024}),\ \Eprint {https://arxiv.org/abs/2405.09629} {arXiv:2405.09629 [hep-ph]} \BibitemShut {NoStop}%
\bibitem [{\citenamefont {Kansal}\ \emph {et~al.}(2021)\citenamefont {Kansal}, \citenamefont {Duarte}, \citenamefont {Su}, \citenamefont {Orzari}, \citenamefont {Tomei}, \citenamefont {Pierini}, \citenamefont {Touranakou}, \citenamefont {Vlimant},\ and\ \citenamefont {Gunopulos}}]{kansal2021mpgan}%
  \BibitemOpen
  \bibfield  {author} {\bibinfo {author} {\bibfnamefont {R.}~\bibnamefont {Kansal}}, \bibinfo {author} {\bibfnamefont {J.}~\bibnamefont {Duarte}}, \bibinfo {author} {\bibfnamefont {H.}~\bibnamefont {Su}}, \bibinfo {author} {\bibfnamefont {B.}~\bibnamefont {Orzari}}, \bibinfo {author} {\bibfnamefont {T.}~\bibnamefont {Tomei}}, \bibinfo {author} {\bibfnamefont {M.}~\bibnamefont {Pierini}}, \bibinfo {author} {\bibfnamefont {M.}~\bibnamefont {Touranakou}}, \bibinfo {author} {\bibfnamefont {J.-R.}\ \bibnamefont {Vlimant}},\ and\ \bibinfo {author} {\bibfnamefont {D.}~\bibnamefont {Gunopulos}},\ }\href@noop {} {\bibinfo {title} {{Particle Cloud Generation with Message Passing Generative Adversarial Networks}}} (\bibinfo {year} {2021}),\ \Eprint {https://arxiv.org/abs/2106.11535} {arXiv:2106.11535 [cs.LG]} \BibitemShut {NoStop}%
\bibitem [{\citenamefont {Käch}\ \emph {et~al.}(2022)\citenamefont {Käch}, \citenamefont {Krücker}, \citenamefont {Melzer-Pellmann}, \citenamefont {Scham}, \citenamefont {Schnake},\ and\ \citenamefont {Verney-Provatas}}]{kach2022jetflow}%
  \BibitemOpen
  \bibfield  {author} {\bibinfo {author} {\bibfnamefont {B.}~\bibnamefont {Käch}}, \bibinfo {author} {\bibfnamefont {D.}~\bibnamefont {Krücker}}, \bibinfo {author} {\bibfnamefont {I.}~\bibnamefont {Melzer-Pellmann}}, \bibinfo {author} {\bibfnamefont {M.}~\bibnamefont {Scham}}, \bibinfo {author} {\bibfnamefont {S.}~\bibnamefont {Schnake}},\ and\ \bibinfo {author} {\bibfnamefont {A.}~\bibnamefont {Verney-Provatas}},\ }\href@noop {} {\bibinfo {title} {{JetFlow: Generating Jets with Conditioned and Mass Constrained Normalising Flows}}} (\bibinfo {year} {2022}),\ \Eprint {https://arxiv.org/abs/2211.13630} {arXiv:2211.13630 [hep-ex]} \BibitemShut {NoStop}%
\bibitem [{\citenamefont {Buhmann}\ \emph {et~al.}(2023{\natexlab{a}})\citenamefont {Buhmann}, \citenamefont {Kasieczka},\ and\ \citenamefont {Thaler}}]{buhmann2023epicgan}%
  \BibitemOpen
  \bibfield  {author} {\bibinfo {author} {\bibfnamefont {E.}~\bibnamefont {Buhmann}}, \bibinfo {author} {\bibfnamefont {G.}~\bibnamefont {Kasieczka}},\ and\ \bibinfo {author} {\bibfnamefont {J.}~\bibnamefont {Thaler}},\ }\bibfield  {title} {\bibinfo {title} {{EPiC-GAN: Equivariant point cloud generation for particle jets}},\ }\href {https://doi.org/10.21468/SciPostPhys.15.4.130} {\bibfield  {journal} {\bibinfo  {journal} {SciPost Physics}\ }\textbf {\bibinfo {volume} {15}},\ \bibinfo {pages} {130} (\bibinfo {year} {2023}{\natexlab{a}})}\BibitemShut {NoStop}%
\bibitem [{\citenamefont {Käch}\ and\ \citenamefont {Melzer-Pellmann}(2023)}]{kach2023attention}%
  \BibitemOpen
  \bibfield  {author} {\bibinfo {author} {\bibfnamefont {B.}~\bibnamefont {Käch}}\ and\ \bibinfo {author} {\bibfnamefont {I.}~\bibnamefont {Melzer-Pellmann}},\ }\href@noop {} {\bibinfo {title} {{Attention to Mean-Fields for Particle Cloud Generation}}} (\bibinfo {year} {2023}),\ \Eprint {https://arxiv.org/abs/2305.15254} {arXiv:2305.15254 [hep-ex]} \BibitemShut {NoStop}%
\bibitem [{\citenamefont {Schnake}\ \emph {et~al.}(2024)\citenamefont {Schnake}, \citenamefont {Krücker},\ and\ \citenamefont {Borras}}]{schnake2024calopointflow2}%
  \BibitemOpen
  \bibfield  {author} {\bibinfo {author} {\bibfnamefont {S.}~\bibnamefont {Schnake}}, \bibinfo {author} {\bibfnamefont {D.}~\bibnamefont {Krücker}},\ and\ \bibinfo {author} {\bibfnamefont {K.}~\bibnamefont {Borras}},\ }\href@noop {} {\bibinfo {title} {{CaloPointFlow II Generating Calorimeter Showers as Point Clouds}}} (\bibinfo {year} {2024}),\ \Eprint {https://arxiv.org/abs/2403.15782} {arXiv:2403.15782 [physics.ins-det]} \BibitemShut {NoStop}%
\bibitem [{\citenamefont {Leigh}\ \emph {et~al.}(2024{\natexlab{a}})\citenamefont {Leigh}, \citenamefont {Sengupta}, \citenamefont {Quétant}, \citenamefont {Raine}, \citenamefont {Zoch},\ and\ \citenamefont {Golling}}]{pcjedi}%
  \BibitemOpen
  \bibfield  {author} {\bibinfo {author} {\bibfnamefont {M.}~\bibnamefont {Leigh}}, \bibinfo {author} {\bibfnamefont {D.}~\bibnamefont {Sengupta}}, \bibinfo {author} {\bibfnamefont {G.}~\bibnamefont {Quétant}}, \bibinfo {author} {\bibfnamefont {J.~A.}\ \bibnamefont {Raine}}, \bibinfo {author} {\bibfnamefont {K.}~\bibnamefont {Zoch}},\ and\ \bibinfo {author} {\bibfnamefont {T.}~\bibnamefont {Golling}},\ }\bibfield  {title} {\bibinfo {title} {{PC-JeDi: Diffusion for particle cloud generation in high energy physics}},\ }\href {https://doi.org/10.21468/SciPostPhys.16.1.018} {\bibfield  {journal} {\bibinfo  {journal} {SciPost Physics}\ }\textbf {\bibinfo {volume} {16}},\ \bibinfo {pages} {018} (\bibinfo {year} {2024}{\natexlab{a}})}\BibitemShut {NoStop}%
\bibitem [{\citenamefont {Leigh}\ \emph {et~al.}(2024{\natexlab{b}})\citenamefont {Leigh}, \citenamefont {Sengupta}, \citenamefont {Raine}, \citenamefont {Quétant},\ and\ \citenamefont {Golling}}]{pcdroid}%
  \BibitemOpen
  \bibfield  {author} {\bibinfo {author} {\bibfnamefont {M.}~\bibnamefont {Leigh}}, \bibinfo {author} {\bibfnamefont {D.}~\bibnamefont {Sengupta}}, \bibinfo {author} {\bibfnamefont {J.~A.}\ \bibnamefont {Raine}}, \bibinfo {author} {\bibfnamefont {G.}~\bibnamefont {Quétant}},\ and\ \bibinfo {author} {\bibfnamefont {T.}~\bibnamefont {Golling}},\ }\bibfield  {title} {\bibinfo {title} {{Faster diffusion model with improved quality for particle cloud generation}},\ }\href {https://doi.org/10.1103/PhysRevD.109.012010} {\bibfield  {journal} {\bibinfo  {journal} {Physical Review D}\ }\textbf {\bibinfo {volume} {109}},\ \bibinfo {pages} {012010} (\bibinfo {year} {2024}{\natexlab{b}})}\BibitemShut {NoStop}%
\bibitem [{\citenamefont {Buhmann}\ \emph {et~al.}(2023{\natexlab{b}})\citenamefont {Buhmann}, \citenamefont {Ewen}, \citenamefont {Faroughy}, \citenamefont {Golling}, \citenamefont {Kasieczka}, \citenamefont {Leigh}, \citenamefont {Quétant}, \citenamefont {Raine}, \citenamefont {Sengupta},\ and\ \citenamefont {Shih}}]{epicly}%
  \BibitemOpen
  \bibfield  {author} {\bibinfo {author} {\bibfnamefont {E.}~\bibnamefont {Buhmann}}, \bibinfo {author} {\bibfnamefont {C.}~\bibnamefont {Ewen}}, \bibinfo {author} {\bibfnamefont {D.~A.}\ \bibnamefont {Faroughy}}, \bibinfo {author} {\bibfnamefont {T.}~\bibnamefont {Golling}}, \bibinfo {author} {\bibfnamefont {G.}~\bibnamefont {Kasieczka}}, \bibinfo {author} {\bibfnamefont {M.}~\bibnamefont {Leigh}}, \bibinfo {author} {\bibfnamefont {G.}~\bibnamefont {Quétant}}, \bibinfo {author} {\bibfnamefont {J.~A.}\ \bibnamefont {Raine}}, \bibinfo {author} {\bibfnamefont {D.}~\bibnamefont {Sengupta}},\ and\ \bibinfo {author} {\bibfnamefont {D.}~\bibnamefont {Shih}},\ }\href@noop {} {\bibinfo {title} {{EPiC-ly Fast Particle Cloud Generation with Flow-Matching and Diffusion}}} (\bibinfo {year} {2023}{\natexlab{b}}),\ \Eprint {https://arxiv.org/abs/2310.00049} {arXiv:2310.00049 [hep-ph]} \BibitemShut {NoStop}%
\bibitem [{\citenamefont {Mikuni}\ \emph {et~al.}(2023)\citenamefont {Mikuni}, \citenamefont {Nachman},\ and\ \citenamefont {Pettee}}]{mikuni2023fpcd}%
  \BibitemOpen
  \bibfield  {author} {\bibinfo {author} {\bibfnamefont {V.}~\bibnamefont {Mikuni}}, \bibinfo {author} {\bibfnamefont {B.}~\bibnamefont {Nachman}},\ and\ \bibinfo {author} {\bibfnamefont {M.}~\bibnamefont {Pettee}},\ }\bibfield  {title} {\bibinfo {title} {{Fast point cloud generation with diffusion models in high energy physics}},\ }\href {https://doi.org/10.1103/PhysRevD.108.036025} {\bibfield  {journal} {\bibinfo  {journal} {Physical Review D}\ }\textbf {\bibinfo {volume} {108}},\ \bibinfo {pages} {036025} (\bibinfo {year} {2023})}\BibitemShut {NoStop}%
\bibitem [{\citenamefont {Bellagente}\ \emph {et~al.}(2020)\citenamefont {Bellagente}, \citenamefont {Butter}, \citenamefont {Kasieczka}, \citenamefont {Plehn}, \citenamefont {Rousselot}, \citenamefont {Winterhalder}, \citenamefont {Ardizzone},\ and\ \citenamefont {Köthe}}]{bellagente2020invertible}%
  \BibitemOpen
  \bibfield  {author} {\bibinfo {author} {\bibfnamefont {M.}~\bibnamefont {Bellagente}}, \bibinfo {author} {\bibfnamefont {A.}~\bibnamefont {Butter}}, \bibinfo {author} {\bibfnamefont {G.}~\bibnamefont {Kasieczka}}, \bibinfo {author} {\bibfnamefont {T.}~\bibnamefont {Plehn}}, \bibinfo {author} {\bibfnamefont {A.}~\bibnamefont {Rousselot}}, \bibinfo {author} {\bibfnamefont {R.}~\bibnamefont {Winterhalder}}, \bibinfo {author} {\bibfnamefont {L.}~\bibnamefont {Ardizzone}},\ and\ \bibinfo {author} {\bibfnamefont {U.}~\bibnamefont {Köthe}},\ }\bibfield  {title} {\bibinfo {title} {{Invertible networks or partons to detector and back again}},\ }\href {https://doi.org/10.21468/SciPostPhys.9.5.074} {\bibfield  {journal} {\bibinfo  {journal} {SciPost Physics}\ }\textbf {\bibinfo {volume} {9}},\ \bibinfo {pages} {074} (\bibinfo {year} {2020})}\BibitemShut {NoStop}%
\bibitem [{\citenamefont {Howard}\ \emph {et~al.}(2022)\citenamefont {Howard}, \citenamefont {Mandt}, \citenamefont {Whiteson},\ and\ \citenamefont {Yang}}]{howard2022otus}%
  \BibitemOpen
  \bibfield  {author} {\bibinfo {author} {\bibfnamefont {J.~N.}\ \bibnamefont {Howard}}, \bibinfo {author} {\bibfnamefont {S.}~\bibnamefont {Mandt}}, \bibinfo {author} {\bibfnamefont {D.}~\bibnamefont {Whiteson}},\ and\ \bibinfo {author} {\bibfnamefont {Y.}~\bibnamefont {Yang}},\ }\bibfield  {title} {\bibinfo {title} {{Learning to simulate high energy particle collisions from unlabeled data}},\ }\href {https://doi.org/10.1038/s41598-022-10966-7} {\bibfield  {journal} {\bibinfo  {journal} {Scientific Reports}\ }\textbf {\bibinfo {volume} {12}},\ \bibinfo {pages} {7567} (\bibinfo {year} {2022})}\BibitemShut {NoStop}%
\bibitem [{\citenamefont {Quétant}\ \emph {et~al.}(2021)\citenamefont {Quétant}, \citenamefont {Drozdova}, \citenamefont {Kinakh}, \citenamefont {Golling},\ and\ \citenamefont {Voloshynovskiy}}]{turbosim}%
  \BibitemOpen
  \bibfield  {author} {\bibinfo {author} {\bibfnamefont {G.}~\bibnamefont {Quétant}}, \bibinfo {author} {\bibfnamefont {M.}~\bibnamefont {Drozdova}}, \bibinfo {author} {\bibfnamefont {V.}~\bibnamefont {Kinakh}}, \bibinfo {author} {\bibfnamefont {T.}~\bibnamefont {Golling}},\ and\ \bibinfo {author} {\bibfnamefont {S.}~\bibnamefont {Voloshynovskiy}},\ }\href@noop {} {\bibinfo {title} {{Turbo-Sim: a generalised generative model with a physical latent space}}} (\bibinfo {year} {2021}),\ \Eprint {https://arxiv.org/abs/2112.10629} {arXiv:2112.10629 [cs.LG]} \BibitemShut {NoStop}%
\bibitem [{\citenamefont {Quétant}\ \emph {et~al.}(2023)\citenamefont {Quétant}, \citenamefont {Belousov}, \citenamefont {Kinakh},\ and\ \citenamefont {Voloshynovskiy}}]{turbo}%
  \BibitemOpen
  \bibfield  {author} {\bibinfo {author} {\bibfnamefont {G.}~\bibnamefont {Quétant}}, \bibinfo {author} {\bibfnamefont {Y.}~\bibnamefont {Belousov}}, \bibinfo {author} {\bibfnamefont {V.}~\bibnamefont {Kinakh}},\ and\ \bibinfo {author} {\bibfnamefont {S.}~\bibnamefont {Voloshynovskiy}},\ }\bibfield  {title} {\bibinfo {title} {{TURBO: The Swiss Knife of Auto-Encoders}},\ }\href {https://doi.org/10.3390/e25101471} {\bibfield  {journal} {\bibinfo  {journal} {Entropy}\ }\textbf {\bibinfo {volume} {25}},\ \bibinfo {pages} {1471} (\bibinfo {year} {2023})}\BibitemShut {NoStop}%
\bibitem [{\citenamefont {Soybelman}\ \emph {et~al.}(2023)\citenamefont {Soybelman}, \citenamefont {Kakati}, \citenamefont {Heinrich}, \citenamefont {Di~Bello}, \citenamefont {Dreyer}, \citenamefont {Ganguly}, \citenamefont {Gross}, \citenamefont {Kado},\ and\ \citenamefont {Shlomi}}]{soybelman2023setgeneration}%
  \BibitemOpen
  \bibfield  {author} {\bibinfo {author} {\bibfnamefont {N.}~\bibnamefont {Soybelman}}, \bibinfo {author} {\bibfnamefont {N.}~\bibnamefont {Kakati}}, \bibinfo {author} {\bibfnamefont {L.}~\bibnamefont {Heinrich}}, \bibinfo {author} {\bibfnamefont {F.~A.}\ \bibnamefont {Di~Bello}}, \bibinfo {author} {\bibfnamefont {E.}~\bibnamefont {Dreyer}}, \bibinfo {author} {\bibfnamefont {S.}~\bibnamefont {Ganguly}}, \bibinfo {author} {\bibfnamefont {E.}~\bibnamefont {Gross}}, \bibinfo {author} {\bibfnamefont {M.}~\bibnamefont {Kado}},\ and\ \bibinfo {author} {\bibfnamefont {J.}~\bibnamefont {Shlomi}},\ }\bibfield  {title} {\bibinfo {title} {{Set-conditional set generation for particle physics}},\ }\href {https://doi.org/10.1088/2632-2153/ad035b} {\bibfield  {journal} {\bibinfo  {journal} {Machine Learning: Science and Technology}\ }\textbf {\bibinfo {volume} {4}},\ \bibinfo {pages} {045036} (\bibinfo {year} {2023})}\BibitemShut {NoStop}%
\bibitem [{\citenamefont {Butter}\ \emph {et~al.}(2023{\natexlab{a}})\citenamefont {Butter}, \citenamefont {Heimel}, \citenamefont {Hummerich}, \citenamefont {Krebs}, \citenamefont {Plehn}, \citenamefont {Rousselot},\ and\ \citenamefont {Vent}}]{butter2023precision}%
  \BibitemOpen
  \bibfield  {author} {\bibinfo {author} {\bibfnamefont {A.}~\bibnamefont {Butter}}, \bibinfo {author} {\bibfnamefont {T.}~\bibnamefont {Heimel}}, \bibinfo {author} {\bibfnamefont {S.}~\bibnamefont {Hummerich}}, \bibinfo {author} {\bibfnamefont {T.}~\bibnamefont {Krebs}}, \bibinfo {author} {\bibfnamefont {T.}~\bibnamefont {Plehn}}, \bibinfo {author} {\bibfnamefont {A.}~\bibnamefont {Rousselot}},\ and\ \bibinfo {author} {\bibfnamefont {S.}~\bibnamefont {Vent}},\ }\bibfield  {title} {\bibinfo {title} {{Generative networks for precision enthusiasts}},\ }\href {https://doi.org/10.21468/SciPostPhys.14.4.078} {\bibfield  {journal} {\bibinfo  {journal} {SciPost Physics}\ }\textbf {\bibinfo {volume} {14}},\ \bibinfo {pages} {078} (\bibinfo {year} {2023}{\natexlab{a}})}\BibitemShut {NoStop}%
\bibitem [{\citenamefont {Butter}\ \emph {et~al.}(2023{\natexlab{b}})\citenamefont {Butter}, \citenamefont {Huetsch}, \citenamefont {Schweitzer}, \citenamefont {Plehn}, \citenamefont {Sorrenson},\ and\ \citenamefont {Spinner}}]{butter2023jetgpt}%
  \BibitemOpen
  \bibfield  {author} {\bibinfo {author} {\bibfnamefont {A.}~\bibnamefont {Butter}}, \bibinfo {author} {\bibfnamefont {N.}~\bibnamefont {Huetsch}}, \bibinfo {author} {\bibfnamefont {S.~P.}\ \bibnamefont {Schweitzer}}, \bibinfo {author} {\bibfnamefont {T.}~\bibnamefont {Plehn}}, \bibinfo {author} {\bibfnamefont {P.}~\bibnamefont {Sorrenson}},\ and\ \bibinfo {author} {\bibfnamefont {J.}~\bibnamefont {Spinner}},\ }\href@noop {} {\bibinfo {title} {{Jet Diffusion versus JetGPT -- Modern Networks for the LHC}}} (\bibinfo {year} {2023}{\natexlab{b}}),\ \Eprint {https://arxiv.org/abs/2305.10475} {arXiv:2305.10475 [hep-ph]} \BibitemShut {NoStop}%
\bibitem [{\citenamefont {Vaswani}\ \emph {et~al.}(2017)\citenamefont {Vaswani}, \citenamefont {Shazeer}, \citenamefont {Parmar}, \citenamefont {Uszkoreit}, \citenamefont {Jones}, \citenamefont {Gomez}, \citenamefont {Kaiser},\ and\ \citenamefont {Polosukhin}}]{vaswani2023attention}%
  \BibitemOpen
  \bibfield  {author} {\bibinfo {author} {\bibfnamefont {A.}~\bibnamefont {Vaswani}}, \bibinfo {author} {\bibfnamefont {N.}~\bibnamefont {Shazeer}}, \bibinfo {author} {\bibfnamefont {N.}~\bibnamefont {Parmar}}, \bibinfo {author} {\bibfnamefont {J.}~\bibnamefont {Uszkoreit}}, \bibinfo {author} {\bibfnamefont {L.}~\bibnamefont {Jones}}, \bibinfo {author} {\bibfnamefont {A.~N.}\ \bibnamefont {Gomez}}, \bibinfo {author} {\bibfnamefont {L.}~\bibnamefont {Kaiser}},\ and\ \bibinfo {author} {\bibfnamefont {I.}~\bibnamefont {Polosukhin}},\ }\href@noop {} {\bibinfo {title} {{Attention Is All You Need}}} (\bibinfo {year} {2017}),\ \Eprint {https://arxiv.org/abs/1706.03762} {arXiv:1706.03762 [cs.CL]} \BibitemShut {NoStop}%
\bibitem [{\citenamefont {Karras}\ \emph {et~al.}(2022)\citenamefont {Karras}, \citenamefont {Aittala}, \citenamefont {Aila},\ and\ \citenamefont {Laine}}]{karras2022elucidating}%
  \BibitemOpen
  \bibfield  {author} {\bibinfo {author} {\bibfnamefont {T.}~\bibnamefont {Karras}}, \bibinfo {author} {\bibfnamefont {M.}~\bibnamefont {Aittala}}, \bibinfo {author} {\bibfnamefont {T.}~\bibnamefont {Aila}},\ and\ \bibinfo {author} {\bibfnamefont {S.}~\bibnamefont {Laine}},\ }\href@noop {} {\bibinfo {title} {{Elucidating the Design Space of Diffusion-Based Generative Models}}} (\bibinfo {year} {2022}),\ \Eprint {https://arxiv.org/abs/2206.00364} {arXiv:2206.00364 [cs.CV]} \BibitemShut {NoStop}%
\bibitem [{\citenamefont {Durkan}\ \emph {et~al.}(2019)\citenamefont {Durkan}, \citenamefont {Bekasov}, \citenamefont {Murray},\ and\ \citenamefont {Papamakarios}}]{durkan2019neural}%
  \BibitemOpen
  \bibfield  {author} {\bibinfo {author} {\bibfnamefont {C.}~\bibnamefont {Durkan}}, \bibinfo {author} {\bibfnamefont {A.}~\bibnamefont {Bekasov}}, \bibinfo {author} {\bibfnamefont {I.}~\bibnamefont {Murray}},\ and\ \bibinfo {author} {\bibfnamefont {G.}~\bibnamefont {Papamakarios}},\ }\href@noop {} {\bibinfo {title} {{Neural Spline Flows}}} (\bibinfo {year} {2019}),\ \Eprint {https://arxiv.org/abs/1906.04032} {arXiv:1906.04032 [stat.ML]} \BibitemShut {NoStop}%
\bibitem [{\citenamefont {Skands}\ \emph {et~al.}(2014)\citenamefont {Skands}, \citenamefont {Carrazza},\ and\ \citenamefont {Rojo}}]{Monash}%
  \BibitemOpen
  \bibfield  {author} {\bibinfo {author} {\bibfnamefont {P.}~\bibnamefont {Skands}}, \bibinfo {author} {\bibfnamefont {S.}~\bibnamefont {Carrazza}},\ and\ \bibinfo {author} {\bibfnamefont {J.}~\bibnamefont {Rojo}},\ }\bibfield  {title} {\bibinfo {title} {{Tuning PYTHIA 8.1: the Monash 2013 Tune}},\ }\href {https://doi.org/10.1140/epjc/s10052-014-3024-y} {\bibfield  {journal} {\bibinfo  {journal} {The European Physical Journal C}\ }\textbf {\bibinfo {volume} {74}},\ \bibinfo {pages} {3024} (\bibinfo {year} {2014})}\BibitemShut {NoStop}%
\bibitem [{\citenamefont {{NNPDF Collaboration}}\ \emph {et~al.}(2013)\citenamefont {{NNPDF Collaboration}}, \citenamefont {Ball} \emph {et~al.}}]{PartonDFs}%
  \BibitemOpen
  \bibfield  {author} {\bibinfo {author} {\bibnamefont {{NNPDF Collaboration}}}, \bibinfo {author} {\bibfnamefont {R.~D.}\ \bibnamefont {Ball}}, \emph {et~al.},\ }\bibfield  {title} {\bibinfo {title} {{Parton distributions with LHC data}},\ }\href {https://doi.org/10.1016/j.nuclphysb.2012.10.003} {\bibfield  {journal} {\bibinfo  {journal} {Nuclear Physics B}\ }\textbf {\bibinfo {volume} {867}},\ \bibinfo {pages} {244} (\bibinfo {year} {2013})}\BibitemShut {NoStop}%
\bibitem [{\citenamefont {Buckley}\ \emph {et~al.}(2015)\citenamefont {Buckley}, \citenamefont {Ferrando}, \citenamefont {Lloyd}, \citenamefont {Nordström}, \citenamefont {Page}, \citenamefont {Rüfenacht}, \citenamefont {Schönherr},\ and\ \citenamefont {Watt}}]{PartonDAccessLHCC}%
  \BibitemOpen
  \bibfield  {author} {\bibinfo {author} {\bibfnamefont {A.}~\bibnamefont {Buckley}}, \bibinfo {author} {\bibfnamefont {J.}~\bibnamefont {Ferrando}}, \bibinfo {author} {\bibfnamefont {S.}~\bibnamefont {Lloyd}}, \bibinfo {author} {\bibfnamefont {K.}~\bibnamefont {Nordström}}, \bibinfo {author} {\bibfnamefont {B.}~\bibnamefont {Page}}, \bibinfo {author} {\bibfnamefont {M.}~\bibnamefont {Rüfenacht}}, \bibinfo {author} {\bibfnamefont {M.}~\bibnamefont {Schönherr}},\ and\ \bibinfo {author} {\bibfnamefont {G.}~\bibnamefont {Watt}},\ }\bibfield  {title} {\bibinfo {title} {{LHAPDF6: parton density access in the LHC precision era}},\ }\href {https://doi.org/10.1140/epjc/s10052-015-3318-8} {\bibfield  {journal} {\bibinfo  {journal} {The European Physical Journal C}\ }\textbf {\bibinfo {volume} {75}},\ \bibinfo {pages} {132} (\bibinfo {year} {2015})}\BibitemShut {NoStop}%
\bibitem [{\citenamefont {{The ATLAS Collaboration}}(2008)}]{ATLAS}%
  \BibitemOpen
  \bibfield  {author} {\bibinfo {author} {\bibnamefont {{The ATLAS Collaboration}}},\ }\bibfield  {title} {\bibinfo {title} {{The ATLAS Experiment at the CERN Large Hadron Collider}},\ }\href {https://doi.org/10.1088/1748-0221/3/08/S08003} {\bibfield  {journal} {\bibinfo  {journal} {Journal of Instrumentation}\ }\textbf {\bibinfo {volume} {3}},\ \bibinfo {pages} {S08003}}\BibitemShut {NoStop}%
\bibitem [{\citenamefont {Cacciari}\ \emph {et~al.}(2008)\citenamefont {Cacciari}, \citenamefont {Salam},\ and\ \citenamefont {Soyez}}]{AntiKt}%
  \BibitemOpen
  \bibfield  {author} {\bibinfo {author} {\bibfnamefont {M.}~\bibnamefont {Cacciari}}, \bibinfo {author} {\bibfnamefont {G.~P.}\ \bibnamefont {Salam}},\ and\ \bibinfo {author} {\bibfnamefont {G.}~\bibnamefont {Soyez}},\ }\bibfield  {title} {\bibinfo {title} {{The anti-kt jet clustering algorithm}},\ }\href {https://doi.org/10.1088/1126-6708/2008/04/063} {\bibfield  {journal} {\bibinfo  {journal} {Journal of High Energy Physics}\ }\textbf {\bibinfo {volume} {2008}},\ \bibinfo {pages} {063} (\bibinfo {year} {2008})}\BibitemShut {NoStop}%
\bibitem [{\citenamefont {Cacciari}\ \emph {et~al.}(2012)\citenamefont {Cacciari}, \citenamefont {Salam},\ and\ \citenamefont {Soyez}}]{FastJet}%
  \BibitemOpen
  \bibfield  {author} {\bibinfo {author} {\bibfnamefont {M.}~\bibnamefont {Cacciari}}, \bibinfo {author} {\bibfnamefont {G.~P.}\ \bibnamefont {Salam}},\ and\ \bibinfo {author} {\bibfnamefont {G.}~\bibnamefont {Soyez}},\ }\bibfield  {title} {\bibinfo {title} {{FastJet user manual}},\ }\href {https://doi.org/10.1140/epjc/s10052-012-1896-2} {\bibfield  {journal} {\bibinfo  {journal} {The European Physical Journal C}\ }\textbf {\bibinfo {volume} {72}},\ \bibinfo {pages} {1896} (\bibinfo {year} {2012})}\BibitemShut {NoStop}%
\bibitem [{\citenamefont {Shmakov}\ \emph {et~al.}(2024)\citenamefont {Shmakov}, \citenamefont {Greif}, \citenamefont {Fenton}, \citenamefont {Ghosh}, \citenamefont {Baldi},\ and\ \citenamefont {Whiteson}}]{shmakov2024VLD}%
  \BibitemOpen
  \bibfield  {author} {\bibinfo {author} {\bibfnamefont {A.}~\bibnamefont {Shmakov}}, \bibinfo {author} {\bibfnamefont {K.}~\bibnamefont {Greif}}, \bibinfo {author} {\bibfnamefont {M.~J.}\ \bibnamefont {Fenton}}, \bibinfo {author} {\bibfnamefont {A.}~\bibnamefont {Ghosh}}, \bibinfo {author} {\bibfnamefont {P.}~\bibnamefont {Baldi}},\ and\ \bibinfo {author} {\bibfnamefont {D.}~\bibnamefont {Whiteson}},\ }\href@noop {} {\bibinfo {title} {{Full Event Particle-Level Unfolding with Variable-Length Latent Variational Diffusion}}} (\bibinfo {year} {2024}),\ \Eprint {https://arxiv.org/abs/2404.14332} {arXiv:2404.14332 [hep-ex]} \BibitemShut {NoStop}%
\end{thebibliography}%

\clearpage
\appendix

\section{Detailed architecture}
\label{app:architecture_details}

In this appendix we describe the architecture details of the three components of the \pippin model, as well as the main training parameters.%
\footnote{All default hyperparameters are available in the code.}

\subsection{Transformer Encoder}

\begin{figure}[tbh]
    \includegraphics[scale=\diagramscale,trim={0.5cm 0 0.2cm 0},clip]{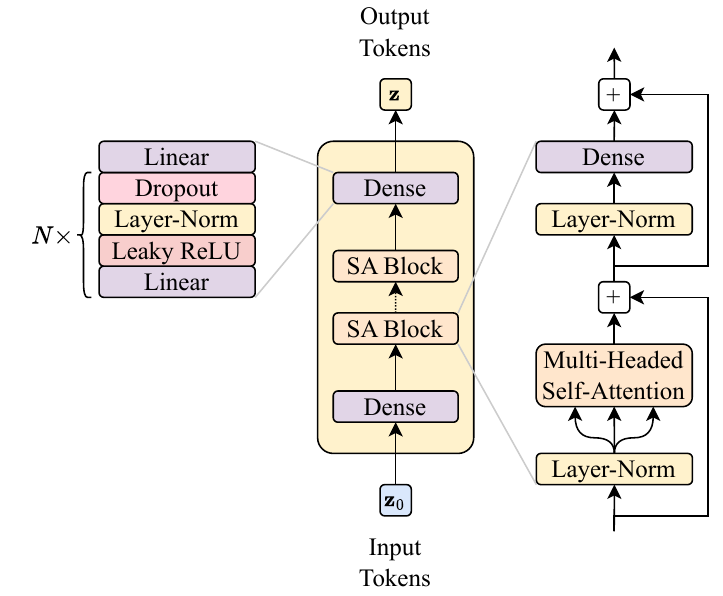}
    \caption{The Transformer Encoder architecture details.}
    \label{fig:transformer_encoder}
\end{figure}

The Transformer Encoder is composed of a node embedding Dense network followed by two Self-Attention (SA) blocks and an output embedding Dense network as shown in \cref{fig:transformer_encoder}.
Each Dense network has a similar architecture, namely stacks of a linear layer, a leaky ReLU activation, a layer-normalisation and a dropout layer.
The SA blocks are the core components of the Transformer Encoder, and are made of a residual Multi-Headed Self-Attention layer and a residual Dense network, both preceded by layer-normalisation.

\subsection{Multiplicity Predictor}

\begin{figure}[tbh]
    \includegraphics[scale=\diagramscale,trim={0.5cm 0 0.1cm 0},clip]{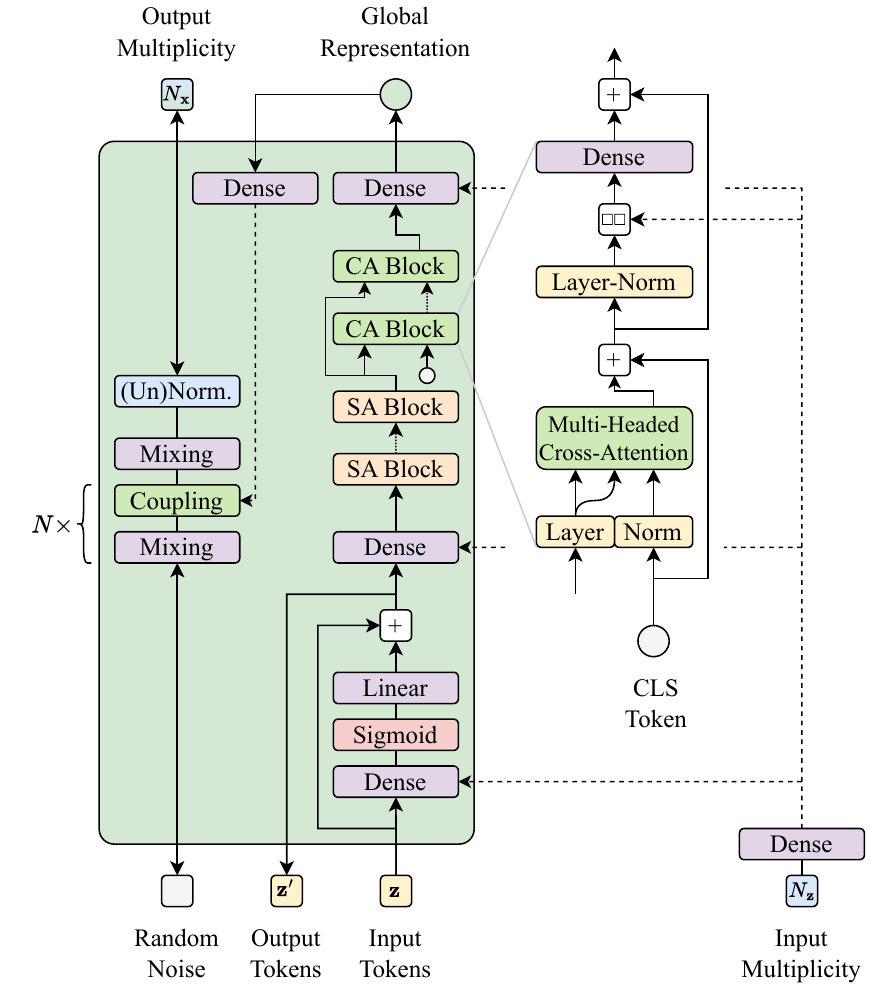}
    \caption{The Multiplicity Predictor architecture details.}
    \label{fig:multiplicity_predictor}
\end{figure}

As shown in \cref{fig:multiplicity_predictor}, the first stage of the Multiplicity Predictor pipeline is a residual presence predictor, conditioned on the input multiplicity, made of a single Dense network whose output is scaled between 0 and 1 by a sigmoid activation and embedded to the input dimension by a linear layer.
The role of this first component is to predict which input partons should be present in the output reconstructed object.
The embedded prediction being residually added to the input, all following layer, including the PIP-Droid Generator, are aware of this presence.

The second stage of the Multiplicity Predictor pipeline is a global representation extractor, also conditioned on the input multiplicity, with an analogue architecture to the Transformer Encoders, but with two additional Cross-Attention (CA) blocks.
These blocks swap the residual Multi-Headed Self-Attention layer for a Cross-Attention version which distributes the input information on a learnable token (CLS) finally representing the extracted global information.

The third stage of the Multiplicity Predictor pipeline is a rational quadratic neural spline coupling normalizing flow.
It is made of stacks of mixing and coupling layers, conditioned on the embedded global representation previously extracted.
The role of this component is the actual prediction of the output reconstructed objects multiplicity.
On top of the leptons and the jets multiplicity, and even though all events contain a single object representing the MET, we decided to let the model predict the MET multiplicity as well in order to better learn how to handle the cases of zero, one and two neutrinos.

\subsection{PIP-Droid Generator}

\begin{figure}[tbh]
    \includegraphics[scale=\diagramscale,trim={0.05cm 0 0 0},clip]{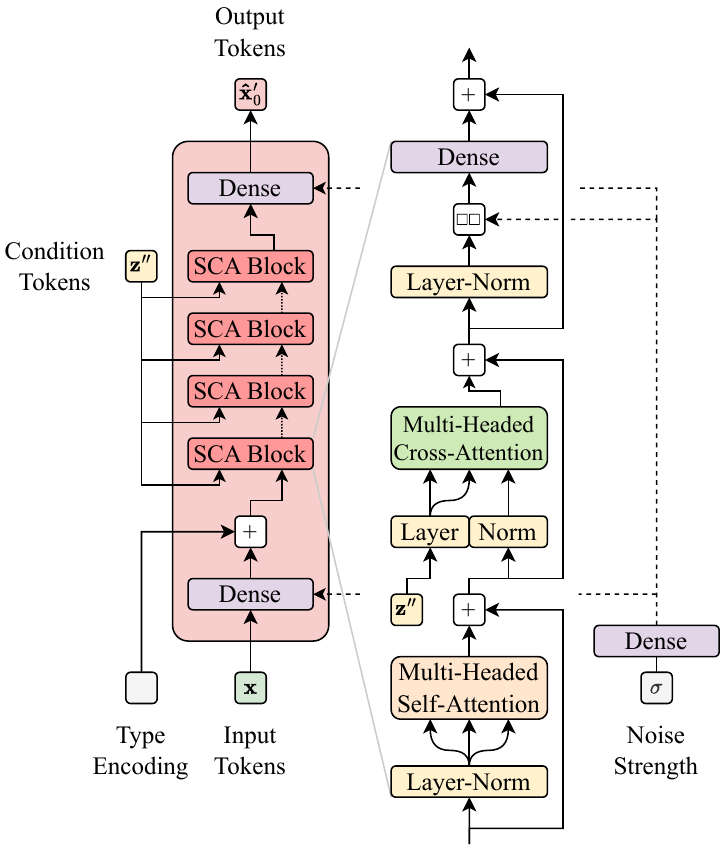}
    \caption{The PIP-Droid Generator architecture details.}
    \label{fig:pipdroid_generator}
\end{figure}

The PIP-Droid Generator has a transformer decoder architecture, shown in \cref{fig:pipdroid_generator}, made of input and output embedding Dense networks, as well as four core Self and Cross-Attention (SCA) blocks, all of them conditioned on the noise strength corresponding to the diffusion step.
Each SCA block contains a first residual Multi-Headed Self-Attention layer to encode the noisy tokens followed by a Multi-Headed Cross-Attention layer to distribute on them the conditional information coming from the parton tokens.
A type encoding learnable token is also added before feeding the SCA blocks.
These tokens are specific to each particle type, namely lepton, MET and jet, in order to avoid the collapse of the output.
Indeed, we observed that the reconstructed objects have a tendency to converge to an average non-physical particle without encoding the desired output type.

\subsection{Training Details}

The whole model is trained end-to-end for 300 epochs with a single combined objective function, except for the first two epochs, for which the Multiplicity Predictor is not used and therefore not trained.
Indeed, giving the true multiplicity at the beginning of the training helps the score-based model to stabilise.
The objective function is composed of a binary cross-entropy loss on the presence prediction, a negative log-likelihood loss on the multiplicity prediction and a mean squared error loss on the output point cloud.
The model is trained with the AdamW optimiser, a batch size of 8192 and a learning rate of $1 \times 10^{-4}$ with a warming up schedule of 50'000 optimisation steps.
We recall that the training dataset contains 37~million events.

Although the global representation extractor and the normalizing flow found in the Multiplicity Predictor form a separate branch of the pipeline, the other branch being the second Transformer Encoder and the PIP-Droid Generator, they are trained together with the rest of the model.
This has two main advantages.
On the one hand, the first few blocks of the \pippin model, namely the first Transformer Encoder and the presence prediction, receive gradient from both branches, since they participate in both.
This helps the convergence of the model by maintaining the consistency of the encoded partons over the Multiplicity Predictor and the PIP-Droid Generator.
On the other hand, it simplifies the training process by avoiding the need to train the two branches separately, which would require a more complex training loop.

\clearpage
\section{Additional plots}
\label{app:additional_plots}

\begin{figure}[H]
    \includegraphics[scale=\plotscale]{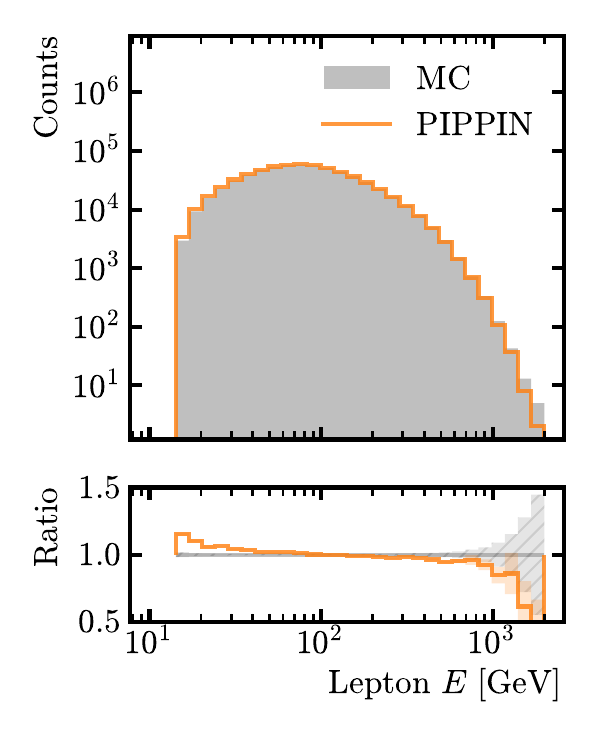}
    \includegraphics[scale=\plotscale]{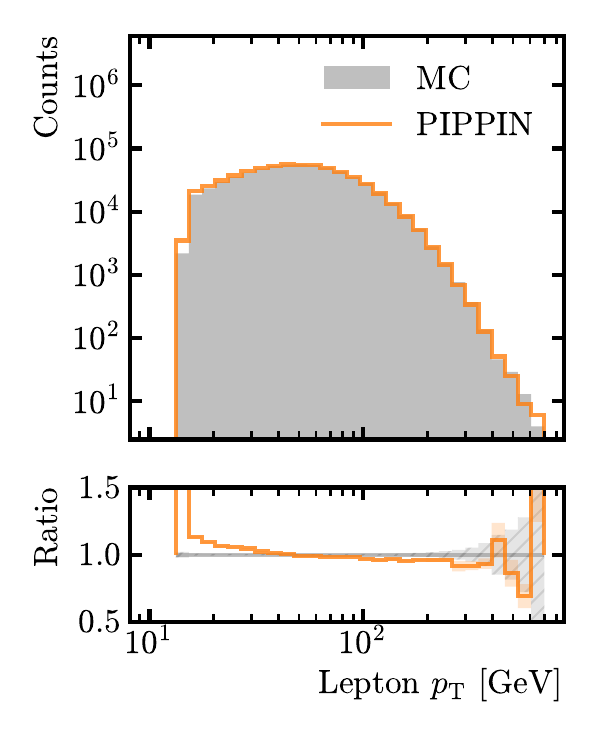}
    \caption{
        Marginal distributions of the learnt features of the reco-level leptons.
        \textbf{Left:} The energy of the leptons in the reconstructed objects.
        \textbf{Right:} The $p_\mathrm{T}$ of the leptons in the reconstructed objects.
        The grey area corresponds to the original MC simulation and the orange line to the output of the \pippin model.
        The bottom plots show the ratios of the histograms with respect to MC and the uncertainties as shaded areas.
    }
    \label{fig:marginals_lep}
\end{figure}

\begin{figure}[H]
    \includegraphics[scale=\plotscale]{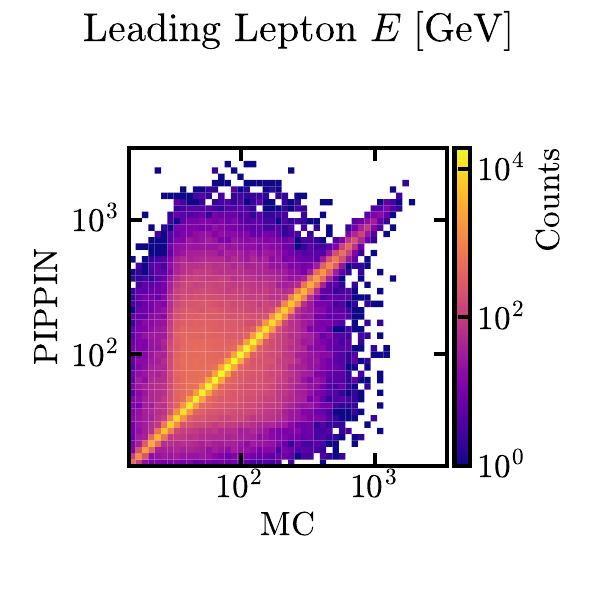}
    \includegraphics[scale=\plotscale]{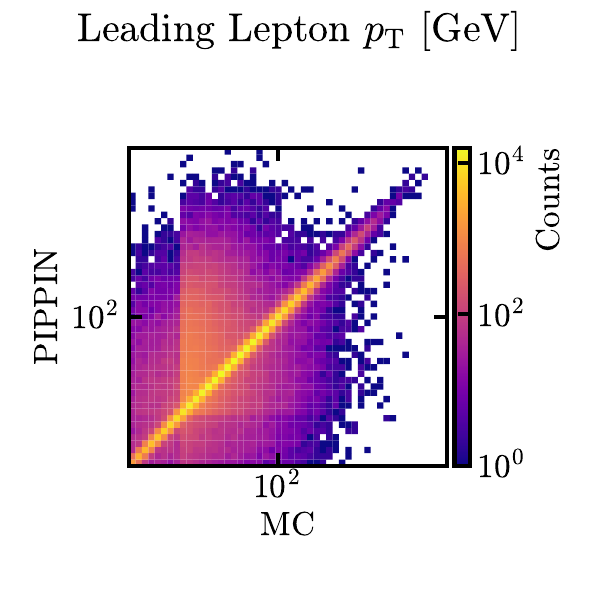}
    \caption{
        2D marginal distributions of the learnt features of the reco-level leptons.
        \textbf{Left:} The energy of the leptons in the reconstructed objects.
        \textbf{Right:} The $p_\mathrm{T}$ of the leptons in the reconstructed objects.
        The $x$-axis corresponds to the original MC simulation and the $y$-axis to the associated output of the \pippin model.
    }
    \label{fig:marginals_2D_lep}
\end{figure}

\vspace*{0.55cm}
\begin{figure}[H]
    \includegraphics[scale=\plotscale]{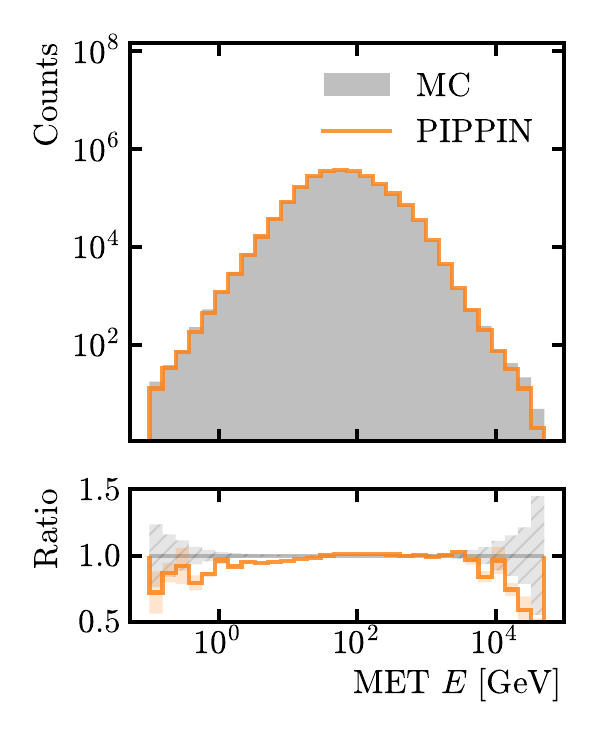}
    \includegraphics[scale=\plotscale]{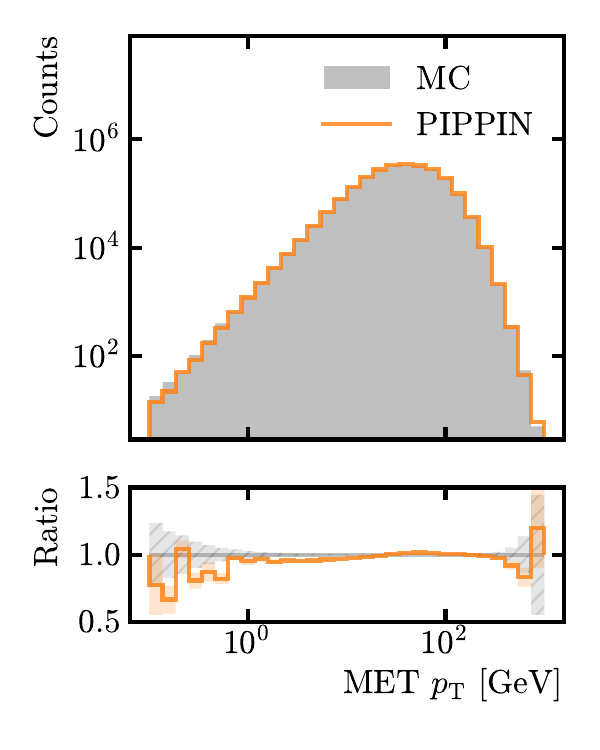}
    \caption{
        Marginal distributions of the learnt features of the reco-level MET.
        \textbf{Left:} The energy of the MET in the reconstructed objects, as calculated in \ref{sec:results_underlying}.
        \textbf{Right:} The $p_\mathrm{T}$ of the MET in the reconstructed objects.
        The grey area corresponds to the original MC simulation and the orange line to the output of the \pippin model.
        The bottom plots show the ratios of the histograms with respect to MC and the uncertainties as shaded areas.
    }
    \label{fig:marginals_met}
\end{figure}

\begin{figure}[H]
    \includegraphics[scale=\plotscale]{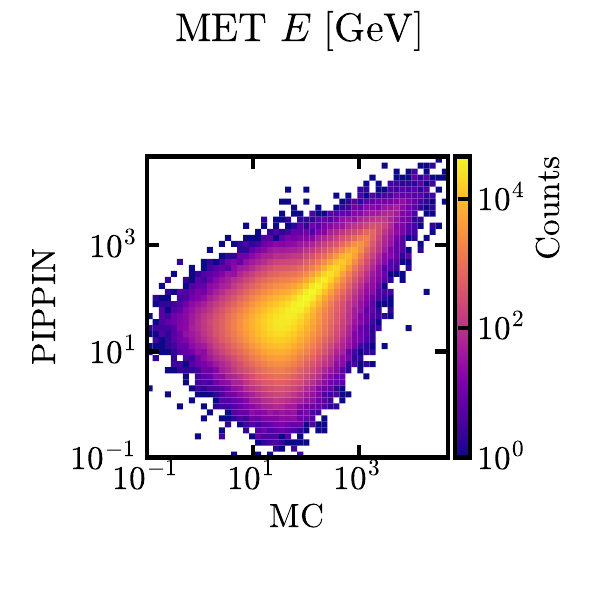}
    \includegraphics[scale=\plotscale]{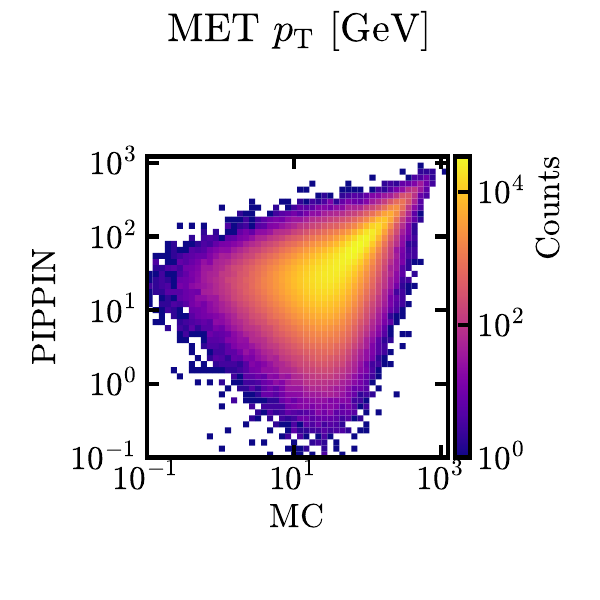}
    \caption{
        2D marginal distributions of the learnt features of the reco-level MET.
        \textbf{Left:} The energy of the MET in the reconstructed objects, as calculated in \ref{sec:results_underlying}.
        \textbf{Right:} The $p_\mathrm{T}$ of the MET in the reconstructed objects.
        The $x$-axis corresponds to the original MC simulation and the $y$-axis to the associated output of the \pippin model.
    }
    \label{fig:marginals_2D_met}
\end{figure}

\begin{figure}[H]
    \includegraphics[scale=\plotscale]{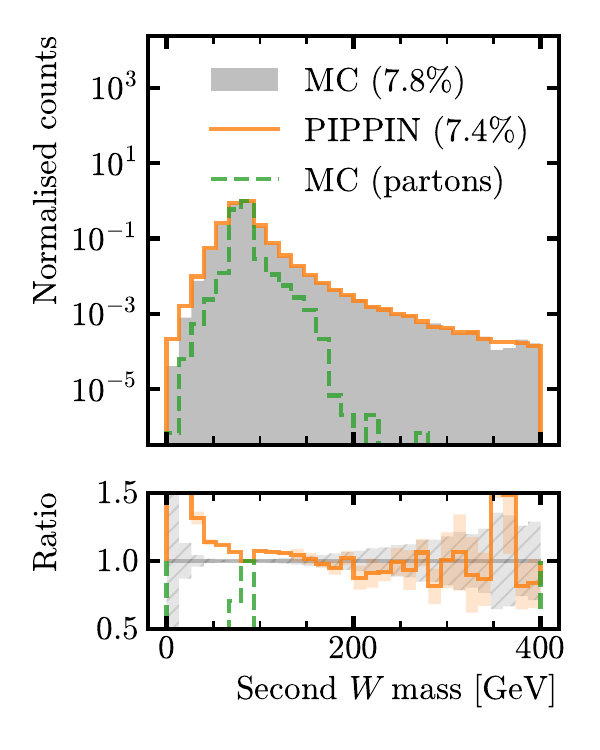}
    \includegraphics[scale=\plotscale]{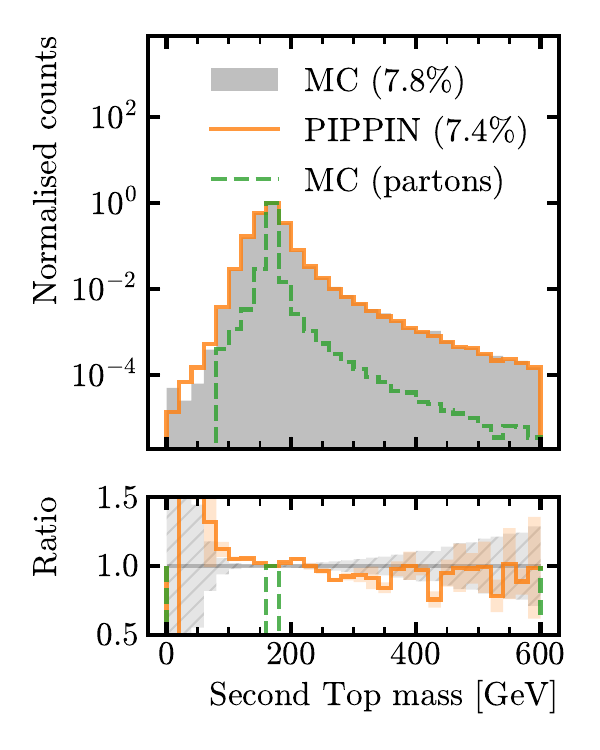}
    \caption{
        Additional marginal distributions of the invariant masses of the underlying particles at reco-level.
        \textbf{Left:} The mass of the second reconstructed $W$ boson.
        \textbf{Right:} The mass of the second reconstructed top quark.
        By second we mean the particle originating from the anti-top quark, as opposed to the top quark.
        The grey area corresponds to the original MC reco-level simulation, the orange line to the output of the \pippin model and the dashed green line to the MC parton-level simulation on which the model is conditioned.
        The bottom plots show the ratios of the histograms with respect to MC and the uncertainties as shaded areas.
        The percentages indicate the proportion of events for which all partons are unambiguously matched and therefore present on the plots.
    }
    \label{fig:masses_additional}
\end{figure}

\begin{figure}[H]
    \includegraphics[scale=\plotscale]{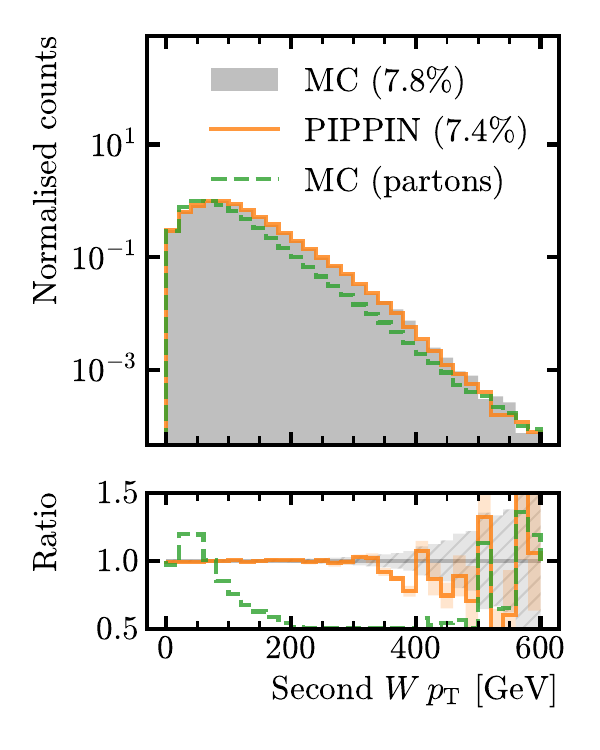}
    \includegraphics[scale=\plotscale]{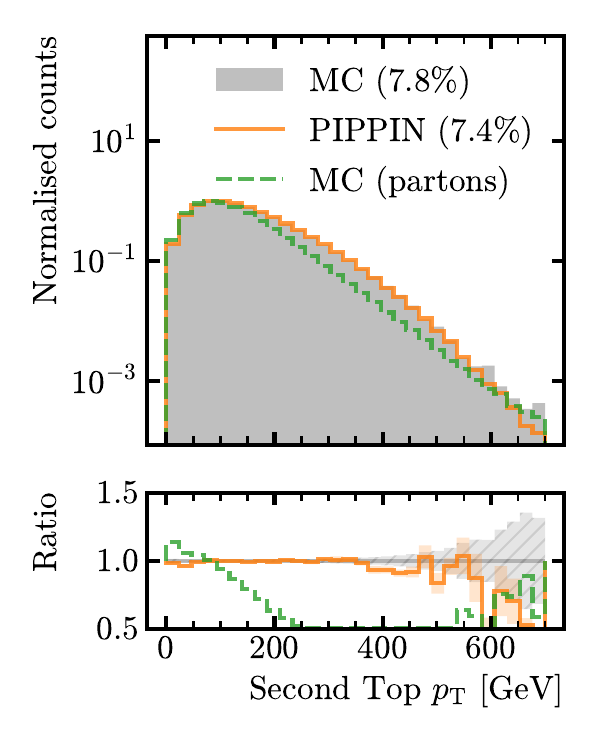}
    \caption{
        Additional marginal distributions of the transverse momenta of the underlying particles at reco-level.
        \textbf{Left:} The transverse momentum of the second reconstructed $W$ boson.
        \textbf{Right:} The transverse momentum of the second reconstructed top quark.
        By second we mean the particle originating from the anti-top quark, as opposed to the top quark.
        The grey area corresponds to the original MC reco-level simulation, the orange line to the output of the \pippin model and the dashed green line to the MC parton-level simulation on which the model is conditioned.
        The bottom plots show the ratios of the histograms with respect to MC and the uncertainties as shaded areas.
        The percentages indicate the proportion of events for which all partons are unambiguously matched and therefore present on the plots.
    }
    \label{fig:momenta_additional}
\end{figure}

\begin{figure}[H]
    \includegraphics[scale=\plotscale]{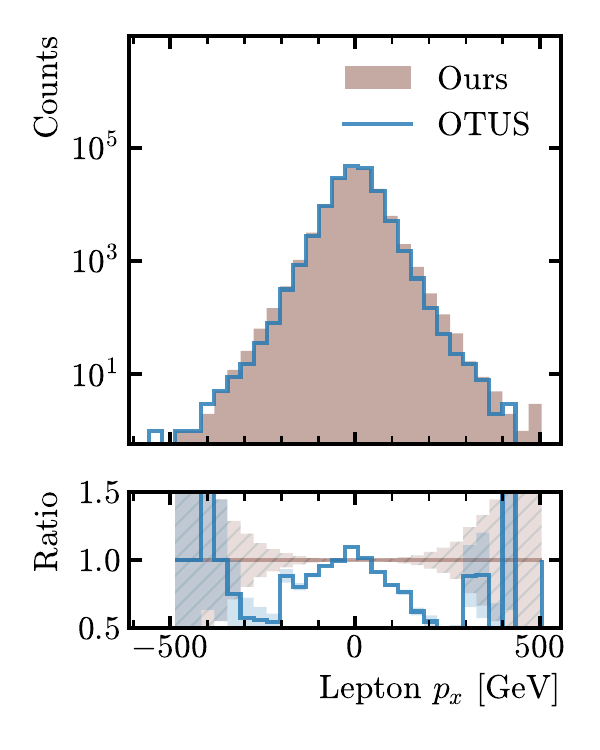}
    \includegraphics[scale=\plotscale]{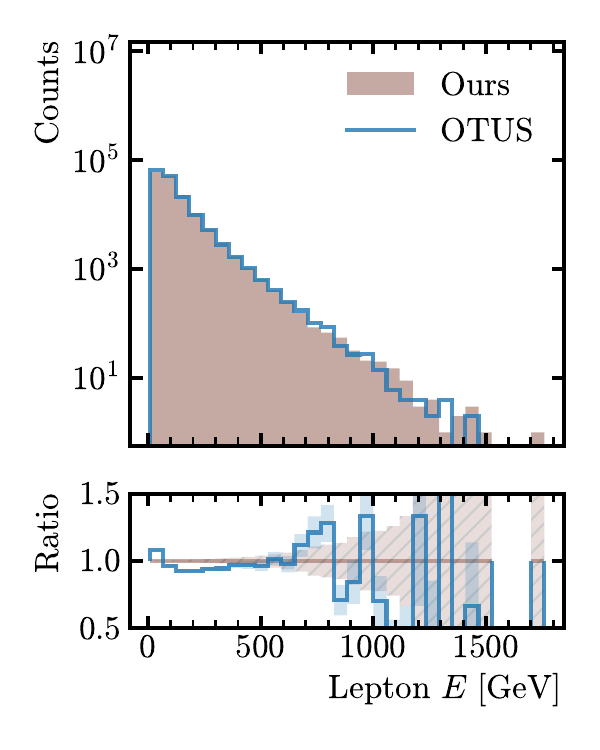}  \\
    \includegraphics[scale=\plotscale]{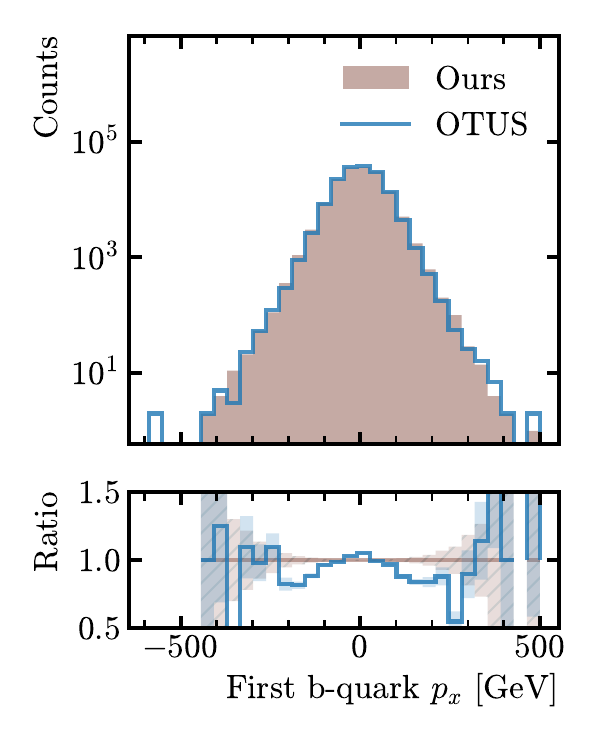}
    \includegraphics[scale=\plotscale]{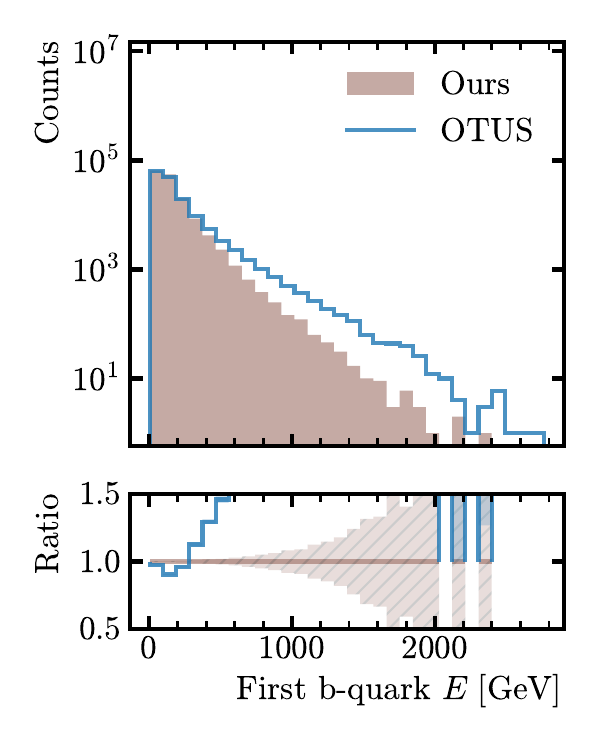}
    \caption{
        Comparison of the parton-level distributions of the lepton and first $b$-quark energy and momentum in the $x$ direction between the dataset presented in this work (brown area) and the one originally used by the OTUS and Turbo-Sim models (blue line).
        The bottom plots show the ratios of the histograms with respect to our dataset and the uncertainties as shaded areas.
    }
    \label{fig:datasets_part}
\end{figure}

\begin{figure}[H]
    \includegraphics[scale=\plotscale]{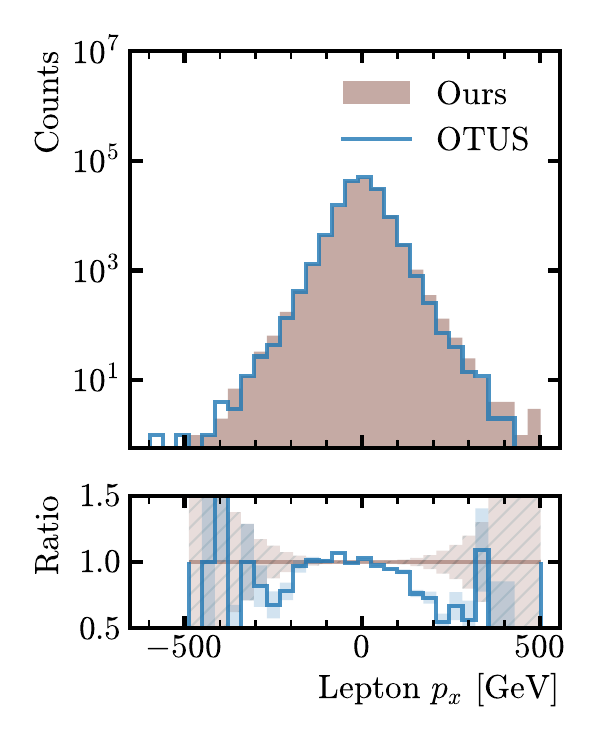}
    \includegraphics[scale=\plotscale]{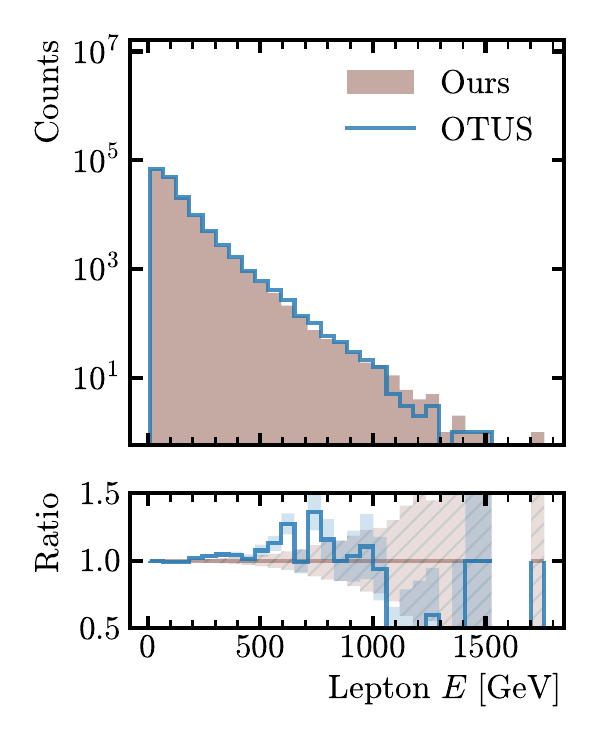} \\
    \includegraphics[scale=\plotscale]{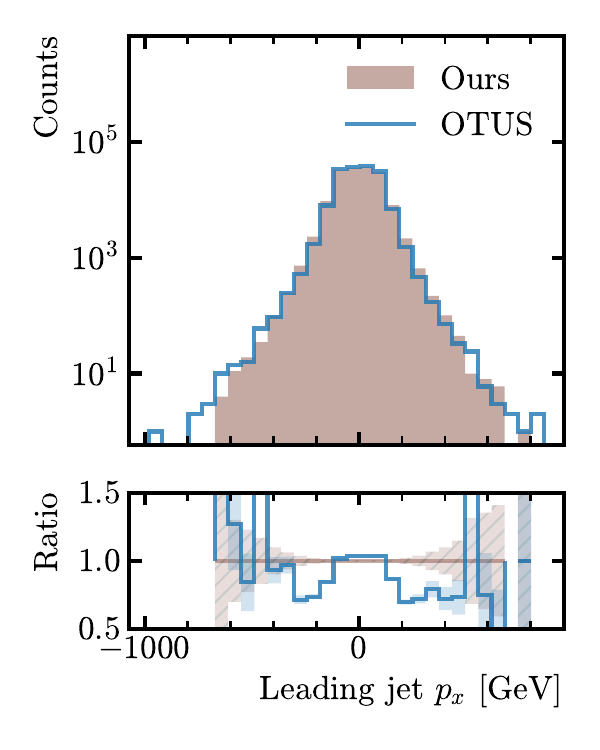}
    \includegraphics[scale=\plotscale]{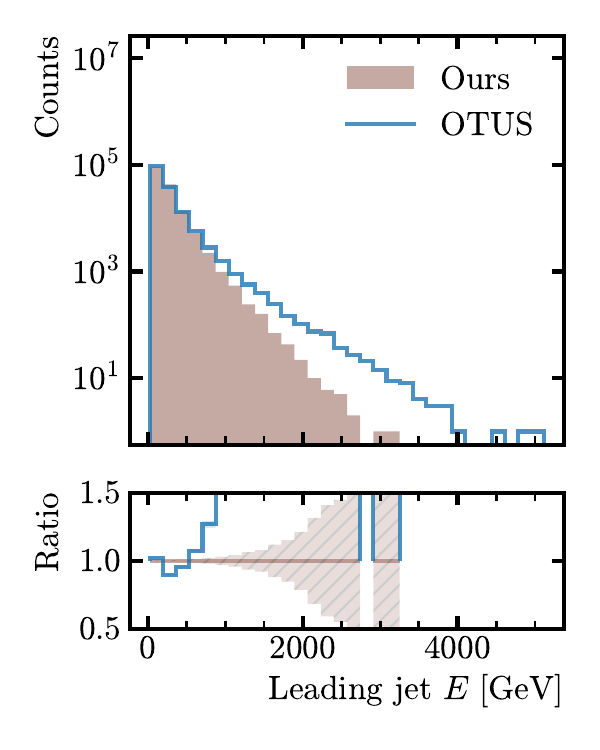}
    \caption{
        Comparison of the reco-level distributions of the lepton and leading jet energy and momentum in the $x$ direction between the dataset presented in this work (brown area) and the one originally used by the OTUS and Turbo-Sim models (blue line).
        The bottom plots show the ratios of the histograms with respect to our dataset and the uncertainties as shaded areas.
    }
    \label{fig:datasets_reco}
\end{figure}

\end{document}